# A Multi-Model Hybrid Defense Approach Against White-box Adversarial Attacks in Computer Network Traffic

Khushnaseeb Roshan*

ORCID ID: https://orcid.org/0000-0002-7114-6168

*Department of Computer Science, Aligarh Muslim University (AMU), Aligarh, INDIA
kroshan@myamu.ac.in / khushiasa@gmail.com

**Abstract**

It is crucial to safeguard computer networks from evolving network security threats and unknown cyberattacks. An essential tool for protecting computer networks against unknown cyber threats is Network Intrusion Detection System (NIDS). However, NIDS faces a major security concern due to its susceptibility to adversarial attacks. Adversarial attacks aim to deceive NIDS by crafting and injecting adversarial examples into the system. These adversarial inputs can deceive the NIDS into misclassifying benign network traffic as malicious. We developed a resilient hybrid defense mechanism aimed to mitigate the impact of two potent adversarial attacks: Fast Gradient Sign Method (FGSM) and Carlini & Wagner (C&W) attack. Our hybrid defense approach leverages the combined strength of two heuristic defense methods: Adversarial Training (AT) and Gaussian Data Augmentation (GDA). GDA provides multi-directional defense, while AT enhances NIDS robustness against specific adversarial vectors. Under pre-attack scenarios, NIDS demonstrated good accuracy and f1-score. However, in the post-attack scenario, its accuracy significantly dropped under FGSM and C&W attacks (0.2649 and 0.4961, respectively). Our proposed hybrid defense method effectively mitigated these adversarial threats, with post-defense accuracy of 96.57% and 89.20% for FGSM and C&W attacks. We evaluated the defense strategy across a range of epsilon and confidence noise factor values (ranging from 0.0001 to 0.0009). This research provides a good direction for future researchers in the emerging area of adversarial machine learning from a security perspective.

## 1. Introduction

Network Intrusion Detection Systems (NIDS) are important tools that help secure computer networks by detecting suspicious activities and potential attacks within the network traffic. It safeguards computer networks against potential security threats and cyberattacks [1], [2], [3]. The primary objective of NIDS is to detect and respond to security events within network traffic to preempt any possible damage. Signature-based and anomaly-based NIDS are the two primary types of NIDS. Signature-based NIDS identifies previously known network intrusions, whereas anomaly-based NIDS identify unknown network attacks. Researchers are dedicating efforts to enhance NIDS performance across various key performance metrics, such as accuracy, F1-score, AUCROC, etc [4], [5], [6]. However, as network security constantly evolves, the capacity for generalization and robustness within NIDS is gaining significant attention. The vulnerability of NIDS to adversarial attack has been shown by recent advancements in adversarial machine learning. The tiny noise is added to real network traffic data to create adversarial examples. These examples have the potential to fool NIDS. It causes them to mistakenly classify benign traffic as malicious and benign traffic as attack.

Adversarial machine learning bridges two broad research domains: computer network security and machine learning. White-box and black-box attacks are the two primary categories into which adversarial attacks are typically divided. White-box adversarial attacks provide the attacker complete access to the target model's internal details, such as its architecture, gradient information, and hyperparameters. On the other hand, black-box attacks presume little or no prior understanding of the internal operations of the model. The other two categories are targeted and untargeted attacks. Targeted attacks involve the attacker manipulating the NIDS model to categorise traffic data into a target class by creating adversarial perturbation input. For instance, benign network traffic can be categorised under a certain attack type, such DDoS, DoS, or PortScan. However, under untargeted attacks, the goal is to reduce the NIDS classification score, such as accuracy, precision, recall, etc. In this study, we consider white-box attacks, where the adversary has full access to the model's architecture and gradients. This setting is commonly adopted in adversarial machine learning research as it represents a worst-case scenario. A model that is robust under white-box conditions is generally expected to perform well against weaker black-box or gray-box threats. Hence enhancing its practical resilience.

In adversarial defense strategy, there are two primary categories: heuristic and certified defense mechanisms [7]. Heuristic defense includes input preprocessing, outlier detection techniques for adversarial noise detection, adversarial training, feature squeezing, data augmentation techniques, etc. Certified defense mechanisms offer formal scientific proof for the model's resilience against such attacks. While these approaches are effective in identifying well-characterized attacks, they often face challenges when encountering novel or unforeseen threats. Heuristic defense methods, on the other hand, are valued for their adaptability in addressing unknown adversarial scenarios, their robustness against evolving evasion techniques, and their implementation flexibility [7]. The defense framework proposed in this study incorporates two heuristic strategies: Adversarial Training (AT) and Gaussian Data Augmentation (GDA). Despite its effectiveness against known threats, adversarial training does not generalize effectively to new attacks. On the other hand, the gaussian data augmentation introduced randomness during the training procedure and is more resilient against evolving adversarial evasion strategies. The proposed novel defense method combined both strategies by using a deep heterogeneous stacking integration method [8], [9] u.

The deep heterogeneous stacking integrated method uses meta-learner technique. In this approach, the output predictions of base models are used as input features for the next model (meta-learner), which is an additional machine learning classifier [10], [11], [12] [13]. However, the dataset's target values (label) remain the same. This way, the overall prediction results improved due to leveraging the strength of each model and capturing different features of the dataset. The same is utilized to increase the defensive capability of the NIDS model.

Fast Gradient Sign Method (FGSM) [14] and Carlini & Wagner (C&W) [15], two well-known adversarial evasion methods, are used to evaluate the efficacy of the suggested hybrid defense approach. The study selects AT, GDA, FGSM, and C&W due to their wide acceptance and representative nature in adversarial machine learning. FGSM offers a fast and standard baseline attack, while C&W represents a strong, optimization-based threat. AT is a robust and widely used defense method against known attacks. GDA enhances model generalization by augmenting data diversity. Together, they provide a balanced evaluation of the proposed method's robustness.

This defense framework integrates two models: AT_NIDS and GDA_NIDS. The AT_NIDS model is developed using adversarial training, incorporating perturbed inputs generated through FGSM and C&W techniques. In parallel, the GDA_NIDS model is trained on Gaussian-augmented data to increase the NIDS resilience. The overall defense mechanism is thoroughly evaluated using classification metrics and confusion matrix under the post-defense phase.

The novelty of the proposed defense method lies in its dual-layered protection. It utilises both AT and GDA. Each method has strengths and weaknesses, but when we use both methods together, it makes a stronger defense mechanism against powerful adversarial attacks. AT enhances robustness to known adversarial inputs, while GDA increases generalization to novel perturbations. Many authors used only single defense methods, such as adversarial training, feature squeezing, gaussian data augmentation, etc. [25], [26]. Furthermore, unlike traditional ensemble methods, our model does not simply average outputs; instead, it trains a secondary classifier on the predictions of both base models to make more informed decisions. To the best of our knowledge, this specific combination of AT and GDA using a meta-learning framework has not been applied to enhance NIDS security under adversarial conditions.

This is the first time we explore the combined effect of adversarial training and Gaussian data augmentation in the context of NIDS, with a focus on robustness under varying perturbation levels.

### *1.1. Novel Contribution*

The novel contributions of this study are presented below:

- Development and evaluation of the NIDS using the CICIDS-2017 benchmark intrusion detection dataset under various performance metrics.
- Comprehensive evaluation and analysis of the NIDS under FGSM and C&W, two powerful adversarial attacks with different epsilon and confidence values (noise factor) varying from 0.0001 to 0.0009.
- Development of the novel hybrid defense method that combined adversarial trained model (AT_NIDS) and gaussian data augmentation model (GDA_NIDS) using meta classifier stacking integrated method.
- Comprehensive analysis to evaluate robustness and resilience of the NIDS after proposed defense to encounter FGSM and C&W adversarial attacks.
- We have included all the technical details, including experiment setups, python libraries, and clear, step-by-step algorithms. Hence, it is easy for future researchers to further enhance the proposed work.

### *1.2. Paper Structure*

An overview of NIDS, adversarial attacks, and defense strategies is given in Section 2, which is followed by a survey of relevant literature. Section 3 consists detailed explanation of the CICIDS-2017 dataset. In Section 4, we describe our methodology, which encompasses the design of adversarial attacks and defense method followed by the proposed hybrid defense method. Section 5 outlines our experimental setup and the libraries/tools used to facilitate transparency and reproducibility in our research. In Section 6, the research findings and outcomes are detailed and analyzed comprehensively. The limitations are described in Section 7, along with suggestions for further investigation. Section 8 presents the concluding remarks of the proposed research.

## 2. Background and Literature Review

This section discusses the machine learning-based NIDS, various terminologies related to adversarial attacks and defense methodologies. It is important to briefly explain these terms to the proposed research study to understand its core concept. In literature review, we have discussed the brief background of adversarial machine learning, followed by the recent state-of-the-art study of adversarial machine learning in the NIDS perspective with its drawbacks and limitations.

### *2.1. Network Intrusion Detection System*

NIDS are crucial for protecting network traffic from unwanted harm, as they continuously analyze network traffic to identify malicious activities within computer network traffic [16]. These systems are like the security guards of the network, protecting against cyber hackers and various security threats.

A deep learning-based NIDS leverages deep neural networks to detect and respond to network intrusions and malicious activities. A typical NIDS is a complex, multi-layered neural network model trained on the historical network dataset to classify benign and attack traffic. The model is configured using specific hyperparameters, including the number of layers, batch size, and learning rate, to achieve optimal classification performance. Activation functions like the Rectified Linear Unit (ReLU) and Sigmoid are used within the neural network to introduce non-linearity. It allows it to capture and learn complex patterns present in network traffic data. Figure 1 illustrates the standard deployment of a NIDS across a network for analyzing and classifying network traffic.

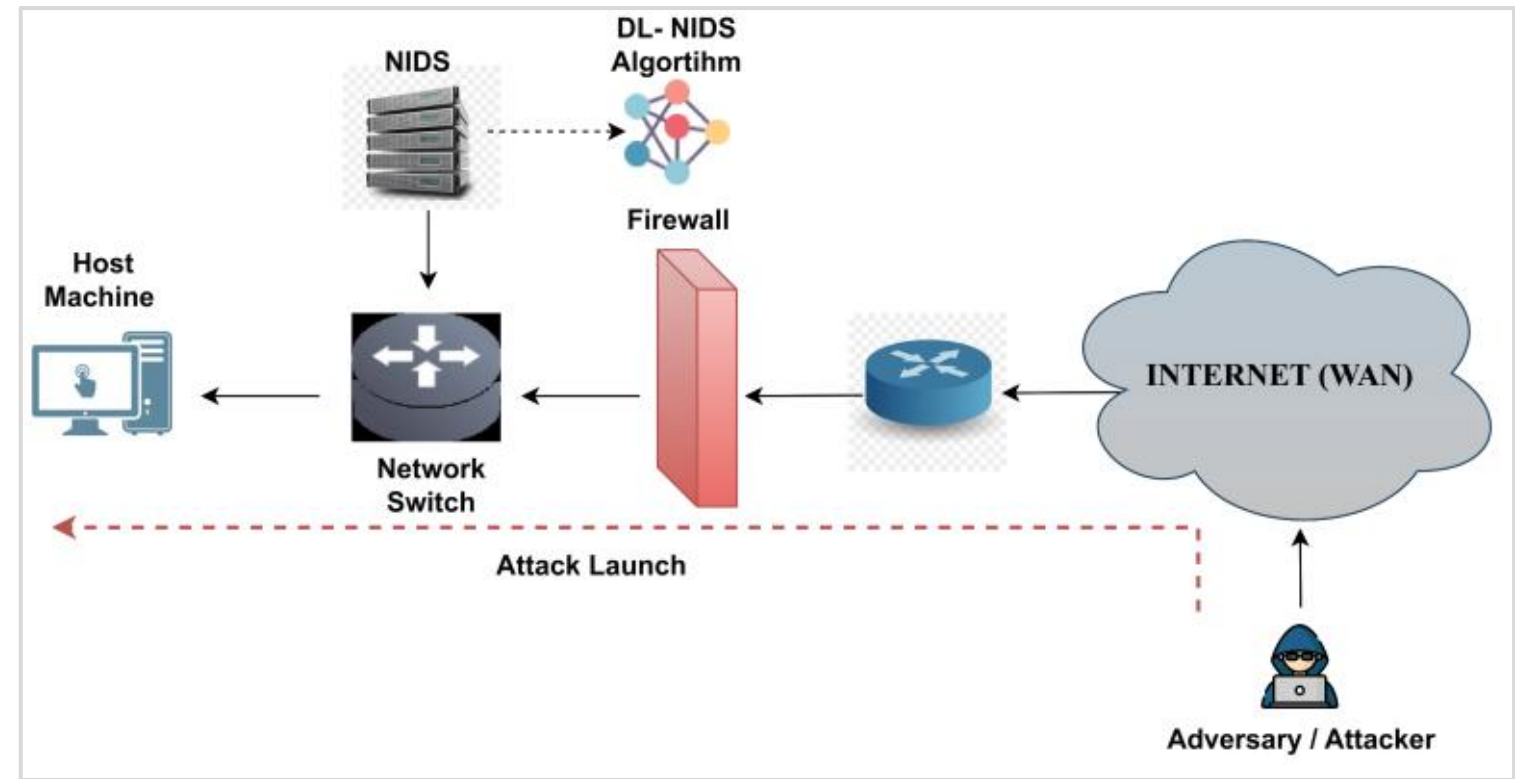

Figure 1 Network-based NIDS Block Architecture with Launching of Adversarial Attack

### *2.2. Adversarial Attacks and Defense Taxonomy*

A simplistic approach is followed to clarify the concept for interested readers. However, we recommend [17] [18] [19] for the detailed technical explanation.

According to the National Institute of Standards and Technology (NIST), adversarial machine learning involves extracting information about the NIDS's behaviour or manipulating inputs to achieve a desired outcome. The field is primarily divided into four adversarial attack types—evasion, poisoning, extraction, and inference which are further discussed in relation to network intrusion detection systems (NIDS). Each attack type targets specific aspects of NIDS. Evasion attacks mislead the model during testing, poisoning attacks compromise the training phase by injecting malicious data, extraction attacks reveal model details and inference attacks exploit prediction confidence to uncover training data. Additionally, these attacks can be classified by timing (training vs. testing), adversary knowledge (white-box vs. black-box), and purpose (targeted manipulation versus untargeted performance degradation). The same approach is also illustrated in our previous research study in detailed.

The adversarial defense methods are categorized into two categories: heuristic defense and certified defense. Heuristic defense relies on practical rules and empirical observations, such as data normalization and adversarial training to quickly adapt to unforeseen threats. Hence making it effective in real-world scenarios where novel attack patterns emerge. Certified defense, on the other hand, utilizes rigorous mathematical modeling and verification techniques, such as Interval Bound Propagation, to provide strong guarantees against specific, well-defined attack patterns. Each approach offers distinct advantages and limitations. Heuristic defenses are valued for their flexibility, speed of implementation, and capacity to detect threats without known signatures, though they may lack formal robustness guarantees. In contrast, certified defenses excel at securing the system against known attacks with mathematical certainty, yet they can struggle with novel or evolving evasion techniques. In our study, we integrate two adversarial defensive methods, AT and GDA to reinforce DL-NIDS against widely recognized adversarial attacks, namely FGSM and C&W, thereby enhancing the system's overall security and robustness. Heuristic defense are flexible, less complex, and contribute to reducing false positives. In contrast, certified defenses are good for known attack signatures and may be susceptible to evasion techniques [20].

We have combined two adversarial defensive methods, namely AT and GDA, to defend DL-NIDS against two powerful attacks, namely FGSM and C&W.

### *2.3. Literature Review*

#### *2.3.1. Adversarial Machine Learning*

Understanding the development and importance of adversarial machine learning in the field of cybersecurity (e.g., NIDS) is essential. In order to increase the robustness and security of machine learning models, adversarial machine learning includes methods for identifying, addressing, and comprehending weaknesses.

Adversarial machine learning started in the early 2000s when researchers noticed that machine learning models, could be tricked by adversarial attacks. One of the earliest instances was observed in spam email classification, where Dalvi et al. (2004) deliberately altered email content to deceive spam filters [21]. This pioneering work shed light on the fact that machine learning models could be easily fooled by subtle changes in input data. Lowd and Meek (2005) developed the concept of Adversarial Classifier Reverse Engineering (ACRE) for reverse engineering adversarial classifiers, laying the foundation for understanding the adversarial vulnerabilities during the training phase [22]. Barreno et al. (2006) further expanded the domain by providing a comprehensive taxonomy of attack types on machine learning systems and discussed defense mechanisms [23]. They raised the fundamental question of whether machine learning could indeed be secure. Szegedy et al. (2013) made a discovery by revealing vulnerabilities in deep learning algorithms, where minor perturbations in images could lead models to confidently make incorrect predictions [24].

Goodfellow et al. [25] made a significant contribution to adversarial machine learning. The research proposed the FGSM for generating adversarial examples by highlighting the system's vulnerability. It also discussed defense mechanisms against adversarial examples. Papernot et al. [26] proposed the concept of transferability phenomena of deep learning models. It says that adversarial examples designed to deceive one machine learning model can successfully fool other models, including those with different architectures or training data.

As illustrated by the data in Figure 2 [27], there has been a notable increase in research publications focused on adversarial machine learning. This surge highlights the rising awareness of the vulnerabilities faced by machine learning models when exposed to adversarial attacks and emphasizes the urgent demand for stronger defense mechanisms. Researchers from a wide range of fields have been consistently working on developing strategies and techniques to improve the security and robustness of machine learning systems against these advancing threats.

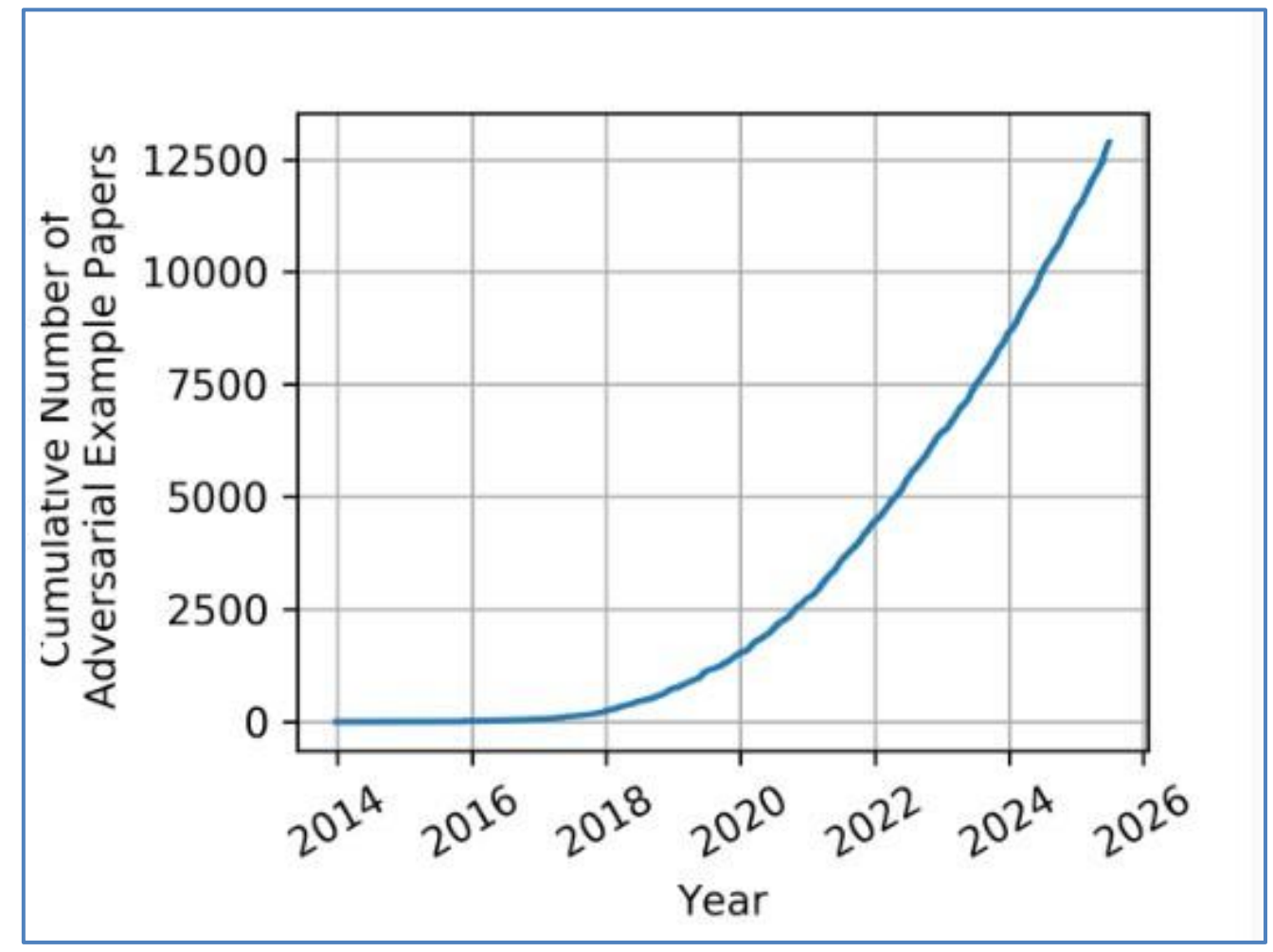

Figure 2 Adversarial Machine Learning Growth Over Years (2014 - 2026) [27]

*2.3.2. Adversarial Machine Learning in NIDS*

In cybersecurity, a key application of adversarial machine learning lies in NIDS, where attackers can take advantage of weaknesses in deep learning models by introducing subtly altered adversarial inputs. The need for generalization and robustness of NIDS becomes increasingly critical. This section discussed the state-of-the-art research in NIDS and adversarial machine learning.

Roshan and Zafar [28] conducted extensive experiments on the CICDDoS-2019 dataset to evaluate the impact of adversarial attacks on cyber attack detection systems. They demonstrated that adversarial perturbations generated using a surrogate model significantly degrade the performance of both surrogate and target models, even in black-box settings. The achieved degraded accuracy for white-box and black-box attacks are 23.85 and 61.15 respectively. The results further reveal that white-box attacks lead to more severe drops in accuracy and F1-score. This emphasized the critical need for robust adversarial defense strategies.

In another research study Roshan and Zafar [29] proposed two-phase defense method against the powerful Carlini & Wagner (C&W) adversarial attack using the CIC-DDoS-2019 dataset. Experimental results showed that the combination of Gaussian Data Augmentation during training and Feature Squeezing (FS) during testing significantly improved the robustness of the NIDS model. The classification reports and confusion matrices confirmed that the defense effectively mitigated the impact of adversarial examples with an improved accuracy of 89.5%.

Sarıkaya et al.[30], tackled the significant challenge posed by adversarial attacks within NIDS. The authors proposed a novel model called RAIDS, which utilized GAN to create realistic adversarial examples and autoencoders to measure differences between input and output data. RAIDS then used an ensemble of machine learning classifiers, including LightGBM, to categorise network traffic as attack or benign. The results demonstrated a significant increase in overall accuracy (by at least 12%) and F1-score (by more than 110%). This approach offers a promising direction for enhancing NIDS robustness without the need for extensive adversarial training.

Debicha et al. [31], addressed the vulnerability of NIDS and emphasized the evasion of adversarial attacks. The authors highlighted the gap between theoretical feasibility and practical implementation of adversarial attacks. The study demonstrated the practicality of fooling NIDS through an adversarial attack implementation. The authors utilized the three machine learning classifiers. The CICIDS-2017 and CTU-13, two benchmark datasets, are used to conduct the study. The authors also implemented the adversarial detector as a defense method to mitigate adversarial attacks. The defense detector improved the robustness of NIDS by using various evaluation metrics.

McCarthy et al. [32], a hierarchical learning-based defense was proposed to counter JSMA attacks. It maintained classification performance comparable to the original NIDS on the CICIDS-2017 dataset. The approach proved effective in a white-box setting, preserving model performance.

Debicha et al. [33] examined NIDS vulnerabilities against FGSM, JSMA, PGD, and DeepFool. They proposed a transfer learning-based detector and showed that using multiple detectors improved adversarial traffic detection, especially in parallel NIDS setups. Experiments on NSL-KDD and CICIDS-2017 highlighted strong performance, though further analysis of detectability versus attack strength was recommended.

Merzouk et al. [34] investigated the practicality of five adversarial attacks, namely, FGSM, BIM, DeepFool, C&W and JSMA, to evade NIDS under white-box attack settings. The authors have done an in-depth analysis of NSL-KDD, UNSW-NB-15, and CIDDS-001, three benchmark network intrusion detection datasets. The detection rate is used as a main evaluation parameter for NIDS under adversarial attack conditions.

Zhang et al. [35] proposed TIKI-TAKA, a comprehensive framework. This study assessed the vulnerability of state-of-the-art NIDS models to various adversarial attack types and highlighted potential success rates of up to 35.7% for attackers. Further, to enhance NIDS resilience, three novel defense mechanisms are proposed: ensembling method, query detection, and voting ensembling. Experimental results demonstrate that these defenses can effectively counter black-box adversarial attacks.

Han et al. [36] explored the vulnerability of NIDS under practical traffic-space attacks, a novel and underexplored area. The author used various machine learning classifiers like KitNET, Isolation Forest, Logistic, Regression, Decision Tree, Multi-Layer Perceptron etc. The key innovation of the proposed metrics lies in the development of an efficient adversarial attack method that can subtly modify network traffic while preserving its functionality. This approach stands out for its practicality, requiring minimal knowledge and computational resources. Furthermore, the proposed defense mechanism demonstrated the attack's effectiveness by 98%.

Maarouf et al. [37] analyzed how well NIDS models can classify encrypted internet traffic when subjected to adversarial attacks. They evaluated five algorithms—C4.5 Decision Tree, KNN, ANN, CNN, and RNN, under adversarial attack scenarios using DeepFool, PGD, and Zoo attacks.

Alhajjar et al. [38] utilized evolutionary methods to generate adversarial examples aimed at evading NIDS. They utilized Particle Swarm Optimization, Genetic Algorithms, and GANs to craft these attacks. The experimentation phase of their study was conducted using two prominent datasets, namely NSL-KDD15 and UNSW-NB15. It enhanced the comprehensiveness of their investigations into adversarial machine learning.

Sethi et al. [39] used context-adaptive IDS using distributed deep reinforcement learning agents like RF, ADB, QDA, GNB, and KNN Over JSMA white-box attack settings. The model is evaluated on three benchmark datasets: AWID, UNSW-NB15, and NSL-KDD. The author also demonstrated robustness against adversarial attacks and used a denoising autoencoder to enhance resilience further.

Pawlicki et al. [40] analyzed the impact of adversarial attacks, namely, FGSM, C&W, PGD, and BIM. They used various machine learning models, including ANN, RF, AdaBoost, and SVM, within the context of network security. Using the CICIDS-2017 dataset, their study highlighted model vulnerabilities and explored strategies to mitigate adversarial threats.

Usama et al. [41] introduced a novel adversarial attack method using Generative Adversarial Networks targeted at black-box-based NIDS. The proposed adversarial attack effectively evaded the NIDS while maintaining the functional integrity of the network traffic features. For evaluation, the study utilizes the KDDCUP-99 benchmark dataset in conjunction with eight machine learning classifiers such as deep neural network, logistic regression, support vector machine etc.

Our novel hybrid algorithm is unique, as none of the previous authors have combined the two defense methods in the way we have combined. By integrating these methods using a meta-heuristics approach, we leverage their strengths to provide a more robust defense against adversarial attacks. Table 1 summarises the related study in NIDS and adversarial machine learning.

Table 1 Related research in adversarial machine learning and NIDS

| **Reference** | **DL Models** | **Dataset** | **Attack Method (Settings)**<br>**WB-White-box,**<br>**BB-Black-box** | **Defense Methods** | **Metrics (%)** | **Key Highlights** |
|---|---|---|---|---|---|---|
| Roshan and Zafar [28] (2024) | DNN Model | CIC-DDoS-2019 | FGSM (WB and BB) | **NA** | Post WB Attack:<br>A: 23.85%,<br>F: 23.0%<br>Post BB Attack: A:61.15%, F:59.94% | - Black-box transferability via surrogate–target models<br>- Real-world CICDDoS-2019 dataset used<br>- Adversarial attack taxonomy included<br>- Only FGSM attack considered |
| Roshan and Zafar [29] (2024) | DNN Model | CIC-DDoS-2019 | C&W (WB) | GDA and FS | Post Defense:<br>A:89.5% F:89.7 %, AUC:89.19 % | - Two-phase defense (Training +Testing) against C&W attack<br>- Evaluated on CIC-DDoS-2019 across five scenarios<br>- Strong post-defense F1-scores (up to 0.895)<br>- Focused only on white-box C&W attack scenario |
| Sarıkaya et al. [30] (2023) | NN, KNN, RAIDS, Light GBM, | CICIDS 2017,<br>InSDN | GAN (WB) | Adversrial Training | Post Defense: A :82.0 %, F1: 90.0% | - Proposed RAIDS: ensemble model using AEs + LightGBM<br>- Used GANs to generate adversarial examples<br>- Achieves +13.2% accuracy, +110% F1 without adversarial training<br>- Limited evaluation on generalization across adversarial attack types |
| Debicha et al. [31] (2023) | MLP, RF, KNN | CSE-CIC-IDS2018 , CTU-13 | Sign Method (BB) | Attack Detector | Detection Rates: 93.4%, and 96.4%, | - Black-box botnet attack using transferability<br>- Model-agnostic ensemble-based reactive defense<br>- Preserved original NIDS performance; good detection results |
| McCarthyet al. [32] (2023) | Ensemble ML, Clustring Algorithm | CICIDS 2017, | JSMA (WB) | Hierarchical Learning | Post Defense:<br>A:90%, F:89% | - Proposed attack-independent hierarchical learning defense<br>- Outperformed under adversarial and clean conditions<br>- Evaluated 7 hierarchy-building methods and improved F1-score<br>- But small sample sizes; needs larger, balanced datasets |
| Debicha et al.<br>[33] (2023) | DNN | NSL-KDD, CICIDS 2017, | FGSM, JSMA, PGD, DeepFool (WB) | Transefer Learning | Pre Attack :A :98% Defense Detection Rate : 89% | - Designed transfer learning-based adversarial detectors<br>- Combined multiple detectors for improved attack detection<br>- Evaluated on state-of-the-art IDS under evasion attacks |
| Merzouk et al. [34] (2022) | MLP | NSL-KDD, UNSW-NB-15, | FGSM, BIM, C&W, JSMA (WB) | NA | Detection Rate: 100%<br>for both | - Evaluated attack impact on NSL-KDD, UNSW-NB15, and CIDDS- |
| | | CIDDS-001 | | | datasets | 001<br>- Proposed syntactic and semantic validity criteria for adversarial examples<br>- Did not proposed a defense<br>- Validity checks were necessary but not sufficient |
| Zhang et al. [35] (2022) | MLP, LSTM, CNN | CICIDS-2017 | PointWise,Opt-Attack, NES, Boundary Attack, HopSkipJumpAttack (BB) | Voting Method, Ensemble Training, Query Method. | Pre Attack: A :99.3 %,<br>Post - Defense: A :99.7% | - Proposed TIKI-TAKA to assess and defend DL-based NIDS<br>- Evaluated 5 attack types on 3 NN-based detectors using public datasets<br>- Introduced 3 defenses: model voting, adversarial training, query detection<br>- Focused on time-based features; broader feature manipulation remained unexplored |
| Han et al. [36] (2021) | SVM, MLP, IF, KitNET, LR, DT, | Kitsune Dataset,<br>CICIDS-2017 | Traffic Mutation<br>(BB<br>GB) | Adversarial feature reduction | Detection :97%, Evasion (MER):97 %<br>Detection: 97% | - Explored GA, and GANs for generating adversarial examples<br>- Demonstrated high evasion rates on NSL-KDD and UNSW-NB15<br>- Evaluated impact on 11 ML models including an ensemble classifier<br>- Focused only on attack strategies; no defense mechanisms were proposed |
| Maarouf et al. [37] (2021) | CNN, DNN, KNN, RNN, C4.5 | SCX-VPN-NON-VPN, NIMS | DeepFool, PGD, Zoo (WB,<br>BB ) | NA | Pre Attack F1: 97%,<br>Post Attack F1 (Min):8% | - Compared ML vs. DL models for encrypted traffic classification<br>- Evaluated resilience |
| | | | | | | under multiple adversarial evasion attacks<br>- Found DL models generally more robust than ML models<br>- DeepFool attack significantly impacted both model types |
| Alhajjar et al. [38] (2021) | BG, LDA, QDA, SVM, KNN, | NSL-KDD, UNSW-NB15 | PSO, GA, GAN (BB) | GAN | Post Attack Evasion Rate: 49.85%<br>Evasion Rate GAN : 99.61 %) | - Explored PSO, GA, and GANs for generating adversarial examples<br>- Demonstrated high evasion rates on NSL-KDD and UNSW-NB15<br>- Evaluated impact on 11 ML models including an ensemble classifier<br>- Focused only on attack strategies; no defense mechanisms were proposed |
| Sethi et al. [39] (2020) | RF, AdaBoost, QDA, GNB, KNN | NSL-KDD, UNSW-NB15, AWID | JSMA (WB) | DAE | Post Defense:<br>A 80.05%, ,<br>AUC:80.01 % | - Proposed context-adaptive IDS using distributed DRL agents<br>- Achieved better accuracy and lower FPR on NSL-KDD, UNSW-NB15, and AWID<br>- Improved robustness with denoising autoencoder and real-world adaptability<br>- Showed limited resistance to adversarial attacks; only minor accuracy drop noted |
| Pawlicki et al. [40] (2020) | ANN, RF, ADABoost, SVM | CICIDS-2017 | FGSM, PGD, BIM,<br>C&W (WB) | Adversarial Detector | Post Defense: P:0.99% R:0.85% | - Used test-time neural activations to detect adversarial attacks<br>- Achieved 0.99 re- |
| | | | | | F:.91% | call using Random Forest and Nearest Neighbour<br>- Evaluated on CICIDS2017 with four major evasion attacks (FGSM, BIM, C&W, PGD)<br>- High false positive rate remained a challenge; limited benchmarking with other domains |
| Usama et al. [41] (2019) | DNN, GAN, K-NN, RF, SVM | KDDCUP-99 | GAN (BB) | Adversarial Training | PreAttack: F: 80.03%,<br>Post Attack: F 40.86%<br>Post Defense: F: 83.56% | - Used GANs to generate adversarial perturbations against ML/DL-based IDSs<br>- Demonstrated successful evasion of IDS without internal model knowledge<br>- Proposed GAN-based defense to improve IDS robustness<br>- Focused evaluation; did not compare with other advanced attack/defense methods |

## 3. Dataset Description

The CICIDS-2017 dataset, developed by the Canadian Institute for Cybersecurity. It, serves as a standard benchmark for evaluating network intrusion detection systems. It is freely accessible for research purposes and comprises traffic data captured in a testbed designed to mimic real-world network conditions and cyberattacks. Although the CICIDS-2017 dataset was released in 2017, it contains sophisticated and recent attack patterns such as DDoS, Botnet, Heartbleed, and Infiltration, which are still considered relevant for modern-day NIDS evaluations. It continues to be widely accepted in recent research as a benchmark dataset for adversarial learning and intrusion detection studies.

The data is organized across five daily files (Monday to Friday), each containing both normal and attack traffic. We use a subset of the CICIDS2017 dataset, including both benign flows and attack types such as DoS, PortScan, Botnet, Web, and Infiltration. All features used in our study are based on flow-level information, such as duration, packet and byte counts, and flow statistics. CICIDS-2017 dataset is relatively large, making it suitable for training and evaluating machine learning models. It includes millions of network flow records. The detailed count-wise descriptions of the dataset are available in [42] [43]. The CICIDS-2017 dataset is chosen for its realistic network traffic and inclusion of modern attack types, making it more representative than older datasets like KDDCUP-99, NSL-KDD, and UNSW-15. A subset of this dataset was used in our study.

### *3.1. Dataset Preprocessing*

The dataset has been preprocessed to evaluate the NIDS model efficiently. The specific preprocessing steps performed over the dataset are illustrated in detail as follows:

Data Cleaning: It is crucial to handle any corrupted and missing value in the dataset for NIDS model accuracy and reliability. In this step, we have checked the dataset for any null data and replaced it with median. We did not choose the mean because as it is suitable for normally distributed data with minimal outliers, as it accurately reflects the central tendency. However, in cases where the data is skewed or contains outliers which is common in real-world network traffic. We used the median, as it is more robust and less sensitive to extreme values. In our study, we primarily applied the median to input missing values. It ensures the consistency and reliability of the feature distribution for training the NIDS model.

Feature Selection and Normalization: We have utilized the feature selection approach from our previous research work [44] [45]. We have utilized kernelSHAP Explainable Artificial Intelligence (XAI) for selecting best feature sets [46].

The study used KernelSHAP to explain the reconstruction error of an autoencoder trained on benign CICIDS2017 data in an unsupervised manner. The explanation model was built using the Value Function by Marginalization with mean squared error as the error metric [47]. Around 200 benign instances as the background set were used to compute Shapley-based feature importance. This provided a more accurate interpretation than raw feature-wise errors. Based on the reconstruction error, the most suitable features were selected. The red colour shows that the particular feature increases the reconstruction error, and the blue colour indicates that this feature reduces the reconstruction error in Figure 3. We have used a total of 44 features, including class labels. The dataset normalization is done with a scikit-learn normalizer using ‘fit’ and ‘transform’ functions.

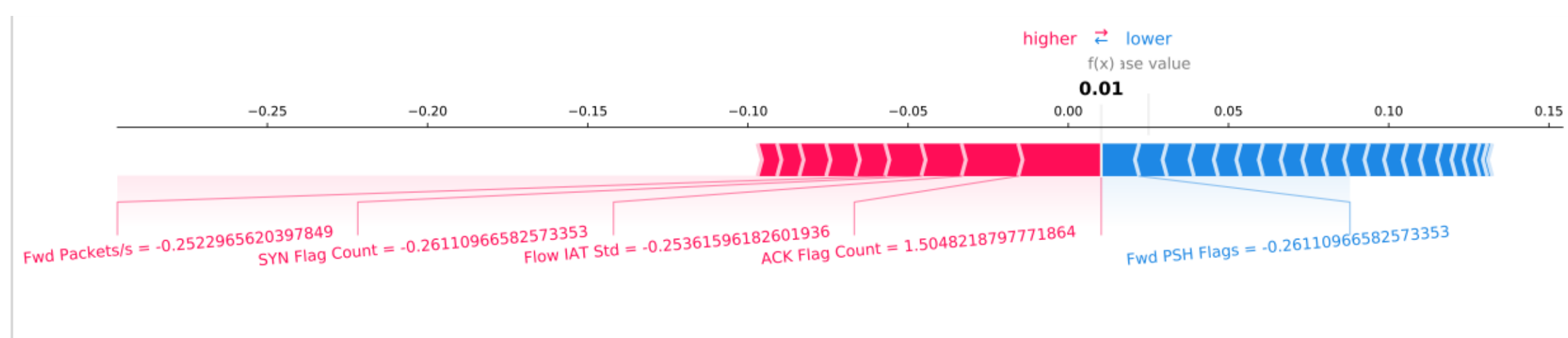

Figure 3 Explainable AI (XAI) Best Features Selection using Model Agnostic Kernel SHAP method to build optimal NIDS model

Data Encoding and Split: Out of the selected 44 features, ‘Label’ feature is categorical, consisting of a class label. We encoded it to binary classification such that ‘0’ represents the Benign/Normal instance and ‘1’ represents the attack instance. The dataset is divided into training and testing sets for the NIDS model. Table 2 represents the instance count used for experimentation purposes. Figure 4 demonstrates the complete step-by-step procedure of dataset preprocessing. Furthermore, we used a subset of the CICIDS-2017 dataset that was carefully selected to minimize class imbalance and ensure fair training. The data was split into 55% training and 45% testing using a fixed random state (42) for reproducibility. As indicated in Table 2, the number of benign and attack samples is relatively balanced in both sets. It allows for unbiased model training and reliable performance evaluation.

Table 2 Dataset Description

| Description | Counts |
|---|---|
| Total dataset samples | 227148 |
| Total benign samples | 107764 |
| Total attack samples | 119384 |
| Training instance count | 124931 |
| Testing instance count | 102217 |
| Training-testing validation split | 45% |
| Random state | 42 |
| Training samples | (‘Normal’, ‘Attack’) – (59100, 65831) |
| Testing samples | (‘Normal’, ‘Attack’) – (48664, 53553) |

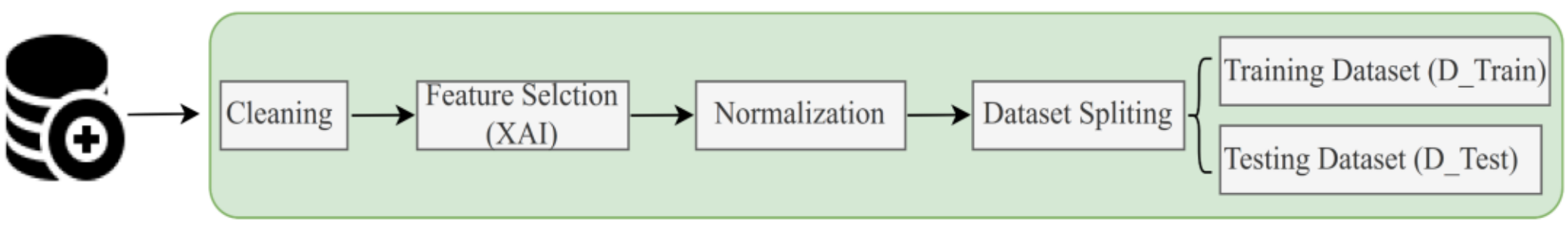

Figure 4 Dataset Preprocessing Steps to Build Optimal NIDS Model

## 4. Methodology

This section describes our methodology for enhancing the resilience and robustness of NIDS through the proposed hybrid defense approach. We start by constructing the NIDS model. Then, we generate adversarial attacks using FGSM and C&W techniques, evaluating NIDS vulnerability. Post-attack, we assess NIDS performance. Our hybrid defense combines adversarial training (AT_NIDS) and Gaussian data augmentation (GDA_NIDS) models. Finally, we evaluate NIDS performance after implementing our defense strategy, aiming to increase its robustness and security. This methodology enables a comprehensive analysis of NIDS performance and security in different scenarios.

### *4.1. Deep Learning based NIDS*

A Deep Neural Network (DNN) is an advanced form of a multilayer perceptron network made up of several layers, including input, hidden, and output layers. Its strength lies in its ability to learn complex patterns from high-dimensional inputs, making it well-suited for identifying intrusions in network traffic data. In the context of NIDS, DNNs are trained to analyze network traffic data and make binary classifications to identify normal (benign) or intrusion (attack) network behaviour. We utilized a subset of the CICIDS-2017 dataset for model evaluation.

In our experiment, the NIDS model consists of one input layer, three hidden layers and one output layer. ReLU is applied between the input and hidden layers for its non-linear properties and efficiency. It helps to address the vanishing gradient issue. Sigmoid function is used in the output layer. A learning rate of 0.0001 ensures stable training, while the Adam optimizer is chosen for its adaptability and strong performance across various models and datasets. The binary_crossentropy is used as the loss function to classify data into benign and attack categories. We addressed overfitting during model training by using cross-validation. Additionally, L2 regularization is applied to prevent overfitting. These techniques ensured that the model generalizes well to unseen data. Table 3 describes the hyperparameter settings utilised for the model building.

Table 3 NIDS Hyperparameter Architecture

| **Hyperparameter** | **Description** | **Values** |
|---|---|---|
| Learning Rate | Step size for weight updates during training | 0.0001 |
| Hidden Layers | Depth of the neural network | 3 |
| Neurons per Layer | Width of each hidden layer | 50,40,20 |
| Input Activation Function | Function applied to neuron outputs for hidden layers | ReLU |
| Output Layer Activation Function | Function applied to neuron outputs for output layer | Sigmoid |
| Batch Size | Number of samples used in each training batch | 1024 |
| Epochs | Number of passes through the entire dataset | 50 |
| Optimizer | Algorithm for updating model weights | Adam |
| Loss Function | Measure of error during training | Binary_crossentropy |
| Regularization | Techniques to prevent overfitting | L2 (0.0001) |
| Split-Parameter | Validation dataset sample to validate model | 0.3 |
| Save_best_only Parameter | To save he best-performing model weights | True |

### *4.2. Adversarial Attack Methods*

#### *4.2.1. Fast Gradient Sign Method (FGSM)*

The Fast Gradient Sign Method (FGSM), introduced by Goodfellow [14], generates adversarial examples by applying minimal, deliberate perturbations to input data to deceive deep learning models. This technique involves calculating the gradient of the loss function with respect to the input and modifying the input in the direction that increases the loss, while keeping the perturbation magnitude constrained using an Lp norm. Specifically, FGSM determines the gradient $\nabla xJ(\Theta, x, y)$, where $\Theta$ denotes the model parameters, x is the input, and y is the true label, and then perturbs x accordingly.

The magnitude of the perturbation, denoted as $\eta$, is controlled by a parameter $\epsilon$. This parameter $\epsilon$ is generally set to a very small value, ensuring that the perturbation is subtle and hard to detect by the human eye but still sufficient to manipulate the model's predictions. By adjusting $\epsilon$, an attacker can control the trade-off between the adversarial sample's stealthiness and its effectiveness in fooling the model. The formulization is illustrated in Equation 1. The core idea of FGSM is to compute the perturbation ($\eta$) to be added to the original input (x) as follows:

$$\eta = \epsilon * sign(\nabla x\, J(\Theta, x, y)) \tag{1}$$

**$\eta$:** The perturbation to be added in data.

**$\epsilon$:** A small scalar value, typically chosen to be a small positive number, e.g., 0.01 or 0.001, depending on the desired level of perturbation.

***sign ($\nabla_x$ J(Θ, x, y)):*** The sign of the gradient of the loss function *J(Θ, x, y*) with respect to the input x. This gradient is calculated using the model's parameters (Θ), the input data (x), and the target output *(y).*

FGSM calculates the loss gradient with respect to the input and introduces a small perturbation in the direction that amplifies the loss, leading the NIDS model to misclassify the altered input.

*4.2.2. Carlini and Wagner (C&W)*

C&W adversarial attack generation technique is presented in [15]. It is a targeted attack method used to create imperceptible perturbations that can deceive DL models. In our study, we have used it as untargeted attacks. i.e., without specifying the target class. The C&W method is based on the optimization-based adversarial attack method. It treats the adversarial attack as an optimization problem, where the goal is to find a perturbation to the input data that minimizes a specific objective function while adhering to certain constraints.

The C&W attack is recognized for its flexibility in applying constraints across different norms—L0, L2, and L∞, to keep perturbations minimal and generally imperceptible to humans. The objective function of the C&W attack is the optimization problem represented by Equation 2:

$$mini\ \delta||\delta||_p + c * f(x + \delta) \tag{2}$$

δ denotes the adversarial noise added to the original input x to create the adversarial example. The term $||\delta||_p$ refers to the norm constraint on this perturbation, where p can represent L0, L2, or L∞ norms. The constant c balances the trade-off between minimizing the perturbation and achieving misclassification. The function f(x + δ) defines the attack objective to be minimized.

Objective Function Choices:

C&W offers multiple choices for the objective function (f(x')). Equation 3 presents one of these choices:

$$f(x') = max(max\ \{Z\ (x')i : i \neq t\} - Z(x')t, -k) \tag{3}$$

Z(*x'*) represents the softmax loss function of the model applied to the adversarial input *x'*.

i ≠ t ensures that the attacker does not target the true class.

k is a confidence constant parameter, required for the model to make incorrect predictions.

In essence, C&W is a highly flexible and powerful targeted adversarial attack that can create imperceptible perturbations while forcing the model to make specific misclassifications. It uses optimization techniques and a variety of objective functions to achieve its goals, making it a notable adversary in the field of machine learning security. Algorithm 1 represents pseudo code to generate adversarial perturbation inputs.

**Algorithm 1**: Pseudocode for generating adversarial inputs using FGSM and C&W methods.

**Input:**

- **NIDS**: Trained NIDS model on the clean dataset
- **D**: Clean dataset to generate adversarial inputs
- **ε (epsilon):** Tiny noise amount for adversarial perturbation for FGSM
- **c (confidence):** Tiny noise amount for adversarial perturbation for C&W

**Output:**

- **D_FGSM**: FGSM Adversarial perturbation examples
- **D_C&W**: C&W Adversarial perturbation examples

---

**Step 1:** FGSM method to generate adversarial perturbations:

a) Create a function called 'FastGradientMethod()' with parameters NIDS (Classifier) and ε (epsilon) values.

```
epsilon_values = {0.0001 to 0.0009}
def FastGradientMethod(NIDS, epsilon_values):
adversarial_examples_FGSM = {}
for c in epsilon_values:
    adversarial_examples_FGSM ← FastGradientSignMethod(NIDS, c)
    adversarial_examples[c] ← adversarial_examples_FGSM
endfor
D_FGSM ← adversarial_examples_FGSM
return D_FGSM
```

**Step 2:** Generate adversarial perturbation using the CarliniL2Wagner (C&W):

a) CarliniL2Method() function with NIDS (Classifier) and c (confidence) values. Apply CarliniL2Method () to the normalized clean dataset D. Generate adversarial perturbation input using confidence (c), noise factor (0.0001 to 0.0009).

```
confidence_values = {0.0001 to 0.0009}
def Carlini&Wagner (NIDS, confidence_values):
adversarial_examples_C&W = {}
for c in confidence_values:
    adversarial_examples_ ← Carlini&Wagner(NIDS, c)
    adversarial_examples[c] ← adversarial_examples_C&W
endfor
D_C&W ← adversarial_examples_C&W
return D_C&W
```

# Both Methods are obtained from the Adversarial-Robustness-Toolbox (ART).

*4.3. Adversarial Defense Method*

In this section, we defined Adversarial Training and Generative Defense Approach for adversarial defense. We then introduced our proposed hybrid defense algorithm, which combined the strengths of both to enhance robustness against adversarial attacks.

*4.3.1. Adversarial Training:*

Adversarial training is a straightforward defense strategy that involves training the model on a mix of clean and adversarially modified inputs. It aims to minimize the loss function not only on clean examples but also on adversarial examples. In the proposed hybrid defense method, we used adversarial training as one of the defense methods. We enrich the training dataset by introducing a subset of generated adversarial examples with FGSM and C&W, thereby strengthening the model's robustness against adversarial attacks that can transfer across different models.

The original NIDS model is represented as 'f(x)', where 'x' is the input and 'y' the corresponding true label. Adversarial training aims to develop a new model 'g(x)' with parameters 'θ_g', designed to resist adversarial perturbations. The objective function for this adversarial training is formulated in Equation 4:

$$argmin\ \theta_g\ [E(x, y) \sim D\ [L(f(x), y) + \alpha * L(g(x_adv), y)]] \quad (4)$$

In this context, 'θ_g' refers to the parameters of the updated NIDS model 'g(x)'. The term 'E(x, y) ~ D' indicates the expected value over the dataset 'D', which includes input samples 'x' and their true labels 'y'. 'L(f(x), y)' represents the loss of the original model 'f(x)' on unaltered data, while 'L(g(x_adv), y)' captures the loss of the new model 'g(x)' when evaluated on adversarial inputs 'x_adv' created using FGSM and C&W. The hyperparameter 'α' balances the emphasis between clean and adversarial loss during training

*4.3.2. Gaussian Data Augmentation:*

GDA technique is commonly used to increase the diversity of the dataset and improve the robustness of the NIDS model. The process involves selecting data points from the original dataset and applying Gaussian noise with a specified mean and standard deviation to each feature dimension. By varying the parameters of the gaussian distribution, different levels of noise can be introduced [20]. The GDA defense strategy offers flexible dataset augmentation by introducing Gaussian-based perturbations. It enables exploration in various directions while encouraging the model to reduce its confidence progressively as inputs deviate from the original sample [20]. The problem is formally expressed as Equation 5:

$$min\theta\ E(x,y) \sim D\ E\Delta x \sim N(0,\sigma 2)\ J(\theta, x + \Delta x, y) \quad (5)$$

In this equation, 'σ' denotes a subtle perturbation level. The goal is to make the posterior distribution 'p(y|x)' align closely with a Gaussian distribution centered at 'x' and having a variance of '$\sigma^2$'.

Subsequently, the authors used a Monte Carlo approximation to address this problem, as depicted in Equation 6:

$$min(\theta)\ E_{(x,y)} \sim D\ [1/N \sum_{i=1}^{n} J(\theta, x + \Delta x_i, y)] \quad (6)$$

In Equation (6), ‘N’ signifies the number of Monte Carlo samples, and ‘$\Delta xi$’ denotes perturbations applied to the input ‘x’ to estimate the expectation over ‘$p(y|x)$.’

In essence, this approach using Gaussian data augmentation offers a flexible strategy for enhancing model robustness. It achieves this by guiding the model away from overconfidence and providing a comprehensive defense against adversarial attacks, ultimately making the model more secure and reliable.

#### *4.3.3. Proposed Hybrid Adversarial Defense*

In our research study, we have proposed a novel defense strategy by combining the strengths of two distinct approaches we discussed above, i.e., GDA trained NIDS model and the AT trained NIDS model. Both approaches form the basis of our proposed hybrid defense method. The proposed defense is also inspired by our previous research work in real-time network streaming data [13]. We would also recommend this work for detailed architecture understanding.

The AT trained model focuses on improving the model’s resistance to adversarial attacks by training on adversarial examples generated through techniques like FGSM and C&W. The GDA trained model, as we discussed earlier, used controlled perturbations following a Gaussian distribution to enhance the model’s robustness and confidence calibration. The GDA defense offers multi-directional defense and gradual confidence reduction, while the AT defense enhances robustness against specific adversarial attack vectors. Through experimental validation and evaluation, we demonstrate the superior performance achieved by the proposed hybrid defense method to safeguard the NIDS against adversarial attacks.

We have combined both the GDA-trained model and the AT-trained model using a stacking integration method. The stacking ensemble method is an ensemble technique that leverages the predictions of multiple base models, in this case, the GDA_NIDS and AT_NIDS, two models combined with the meta-classifier layer that makes the final optimal prediction. Through rigorous experimentation and validation, we demonstrate the effectiveness of our stacking integration method in enhancing the security and reliability of NIDS.

Figure 5 represents the demographic view of the complete proposed methodology, starting from DL_NIDS model building and its pre-attack evaluation. The post-attack evaluation of DL-NIDS on generated adversarial data. Finally, the proposed defense Hybrid_NIDS is explained and evaluated under the post-defense scenario. Algorithm 2 defines the proposed Hybrid Defense Method.

**Algorithm 2:** Hybrid Defense Method using Ensemble Algorithm Pseudo Code

**Input:**

- **AT_NIDS:** Adversarial trained model
- **GDA_NIDS:** Gaussian data augmented trained model
- **D_FGSM:** Adversarial dataset generated using FGSM
- **D_CW:** Adversarial dataset generated using C&W

**Output:**

- **Hybrid_NIDS:** Proposed Hybrid Defense Method
- **Results:** Confusion Matrix, Classification Report, AUC-ROC on D_FGSM and D_CW

---------------------------------------------------------------------------------------------------------------

**Implementation Steps:**

**Step 1:** Predict using AT_NIDS model on D_FGSM and D_CW
predictions_AT_FGSM ← AT_NIDS.**predict**(D_FGSM)
predictions_AT_CW ← AT_NIDS.**predict**(D_CW)

**Step 2:** Predict using GDA_NIDS model on D_FGSM and D_CW
predictions_GDA_FGSM ← GDA_NIDS.**predict**(D_FGSM)
predictions_GDA_CW ← GDA_NIDS.**predict**(D_CW)

**Step 3:** Define a stacked model from the predictions of AT_NIDS and GDA_NIDS.

AT_NIDS.trainable = **false** # freeze the weight and make it non-trainable
GDA_NIDS.trainable = **false** # freeze the weight and make it non-trainable
input_AT ← InputLayer(**shape**=(predictions_AT_FGSM.**shape**[1],))
input_GDA ← InputLayer(**shape**=(predictions_GDA_FGSM.**shape**[1],))

**# concatenate the output of both the model**
concatenated ← Concatenate()([input_AT, input_GDA])

**# add the dense layers with activation functions.**
hidden_layer ← **Dense**(60, activation=**relu**)(concatenated)
output_layer ← **Dense**(2, activation=**sigmoid**)(hidden_layer)
stacked_model ← **Model**(inputs=[input_AT, input_GDA], outputs=output_layer)
stacked_model.**compile**(**optimizer**=adam, **loss**=binary_crossentropy)

**Step 4:** Create the concatenated dataset using the predictions of AT_NIDS and GDA_NIDS models.
concatenated_dataset = pd.**concat**([predictions_AT_FGSM, predictions_GDA_FGSM,
predictions_AT_CW, predictions_GDA_CW], axis=1)

**Step 5**: Fit the stacking model using the concatenated dataset
X_train, X_test, y_train, y_test = **train_test_split**(concatenated_dataset, y, test_size=0.2,
random_state=42)
stacked_model.**fit**([X_train.iloc[:, :2], X_train.iloc[:, 2:]], y_train, epochs=50)

**Step 6:** Evaluate the **Hybrid_NIDS** model on **D_FGSM** and **D_CW**.
y_pred = stacked_model.**predict**([X_test.iloc[:, :2], X_test.iloc[:, 2:]])
confusion_matrix_result = **confusion_matrix**(y_test, y_pred)
classification_report_result = **classification_report**(y_test, y_pred)
auc_roc_result = **roc_auc_score**(y_test, y_pred)

**Step 7:** Return the classification result, confusion metrics, and AUC-ROC.
**return** stacked_model, confusion_matrix_result, classification_report_result, auc_roc_result

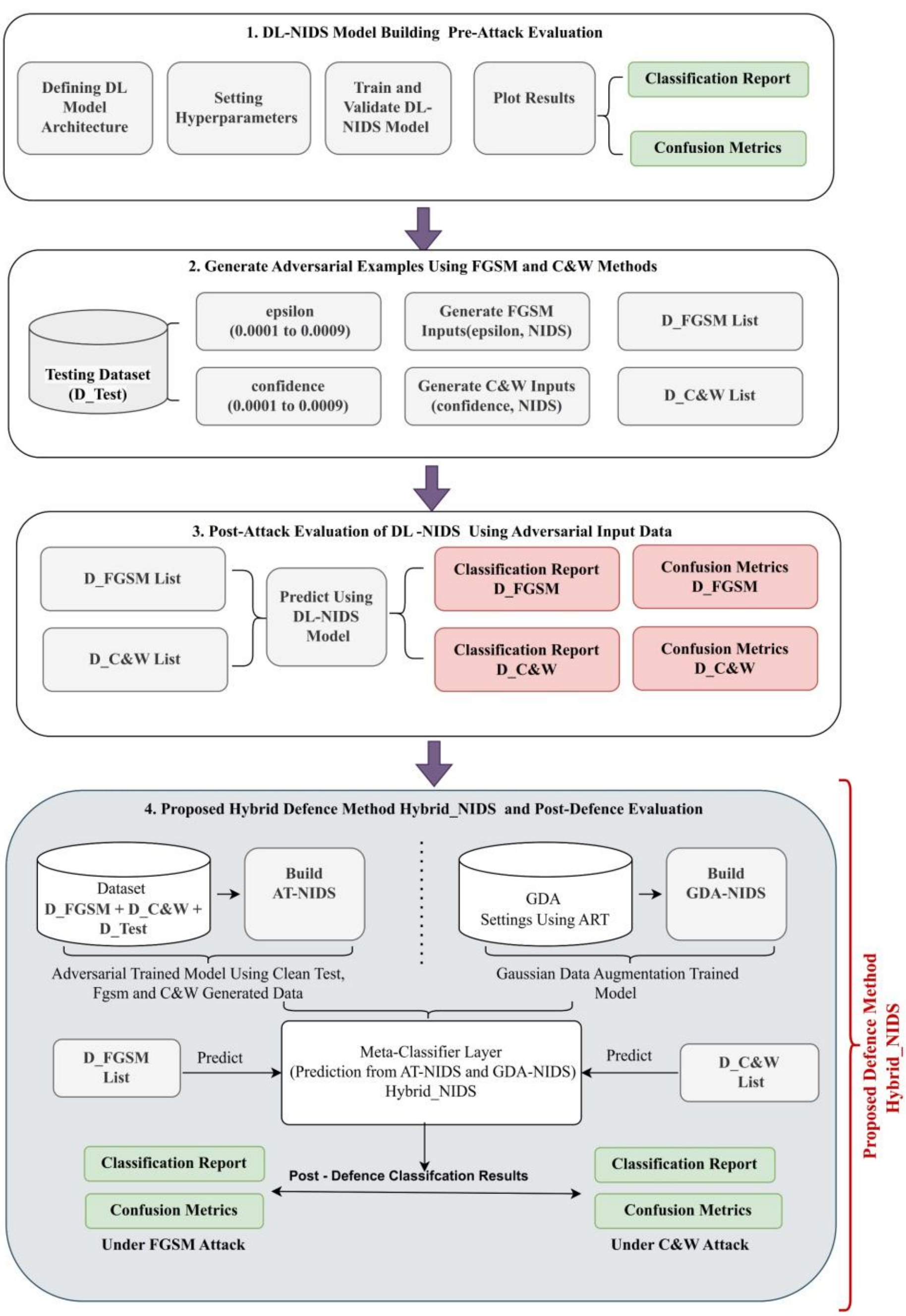

Figure 5 Demographic representation of complete proposed methodology starting from NIDS model building, pre-attack evaluation, post-attack evaluation and post-defense evaluation.

## 5. Experimental Setup

The experiment is carried out on Google Colab, a cloud platform that offers complimentary access to GPU and TPU hardware. It is well-suited for deep learning tasks. It integrates with Google Drive and Jupyter Notebook. It allows to run Python program efficiently. The other supportive libraries required for experimentation are Keras, TensorFlow, Pandas, Numpy, Matplotlib, ART toolbox, etc. The model was trained using Google Colab, with GPU acceleration. The intermediate results and trained models were saved and downloaded for analysis. However, for practical deployment scenarios, the model can be migrated to other platforms such as Jupyter Notebook, Anaconda, or custom cloud/server environments to ensure better integration with real-time NIDS infrastructure and support for large-scale inference.

Keras is a publicly available deep-learning library that offers a user-friendly interface. In our research, we used it to define, compile, and train the DL based NIDS model. Pandas is a powerful data manipulation and analysis library. It allows us to load data from various sources. We used it to load and analyse the CICIDS-2017 dataset for NIDS model building. NumPy provides support for large mathematical calculations required for model building. We used NumPy library to convert the data into a numpy array for NIDS model training and testing. Scikit-learn is a machine-learning library for classification and regression techniques. It provides various evaluation metrics for building machine learning models. We used these libraries for evaluation purpose. The matplotlib is used to demonstrate the results graphically and to plot the NIDS model training plot.

The most important supporting library utilized for the experimentation is the Adversarial Robustness Toolbox (ART). It is a comprehensive library and toolkit specifically designed to evaluate adversarial attack and defense mechanisms using machine learning classifiers. It offers various techniques and functionalities for adversarial defense. The system operated on Windows 11 and featured a hardware setup comprising 16 GB of DDR4L RAM as detailed in Table 4.

Table 4 System Configuration

| **Names** | **Laptop/System Configuration** |
|---|---|
| OS | Windows -11 |
| Memory (RAM) | 16 GB, DDR4L |
| Hard-disk (SSD) | 500 GB |
| CPU | Intel(R)-Core i7,<br>CPU @ 2.20GHz |
| GPU | GeForce GTX GPU, NVIDIA |

## 6. Findings and Discussion

We described a detailed evaluation of the NIDS model across three different scenarios: pre-attack, post-attack, and post-defense. Each scenario highlights the model's performance, adaptability, and robustness when subjected to adversarial attacks, specifically FGSM and C&W methods.

Figure 6(a) and Figure 6(b) illustrate the NIDS model's accuracy and loss on the training and validation datasets, respectively.

The smoothness of the curves indicates that the NIDS model learns the network traffic data efficiently without excessive noise or fluctuations. Hence, it converges towards the optimal solution. The NIDS model is trained for 50 epochs and archives quite acceptable training and validation accuracy, above 95 %. We trained the NIDS model for 50 epochs. This number was selected after observing that both training and validation accuracy began to stabilize around the 45th epoch, as shown in Figure 6(a). Training beyond 50 epochs did not provide significant improvement, and this setting helped avoid overfitting while ensuring convergence.

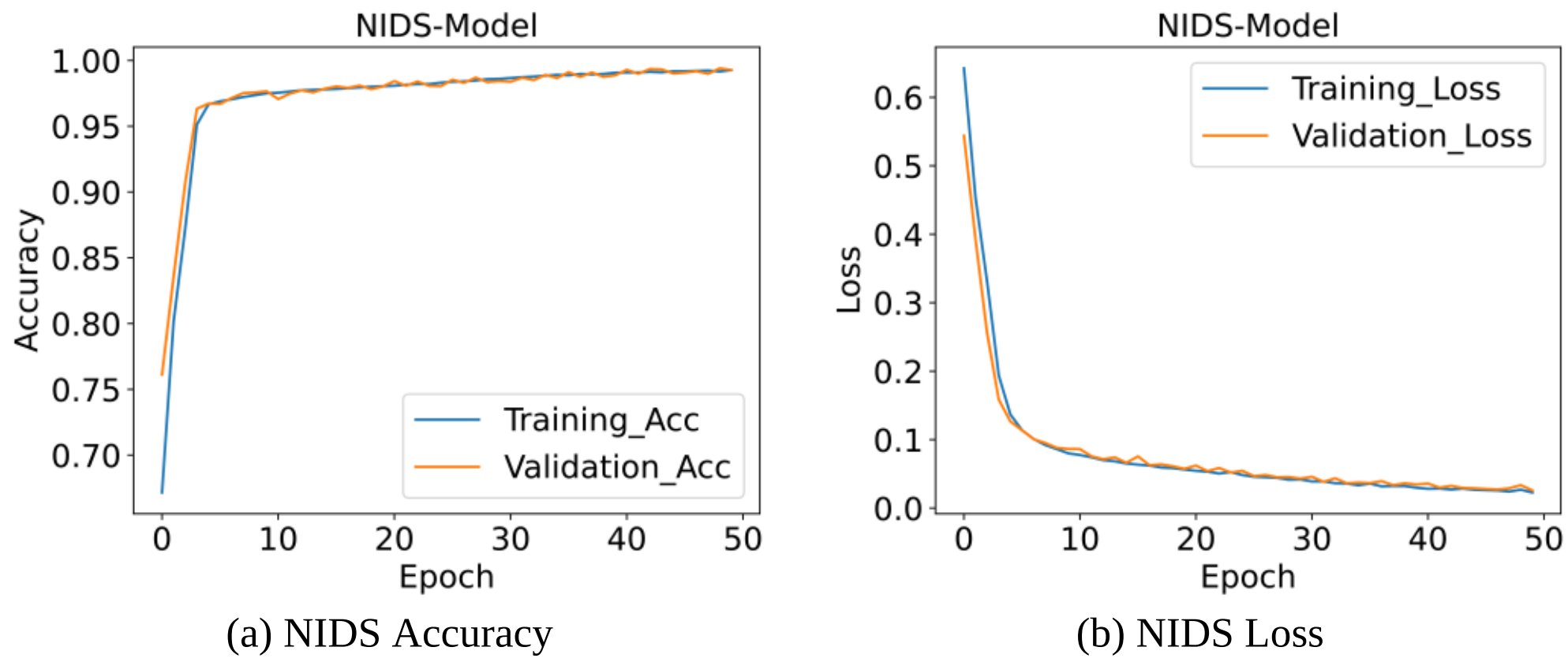

(a) NIDS Accuracy (b) NIDS Loss

Figure 6 (a) Training and validation accuracy (b) Training and validation loss

### *6.1. Performance Before Attack (Pre Attack)*

In a pre-attack scenario, the baseline performance of the NIDS model under standard operating conditions is evaluated. We rigorously tested the model against benign network traffic data, assessing its performance under various performance metrics. This baseline performance provides essential information into the NIDS's capability to identify and classify network intrusions when not subjected to adversarial attacks, such as FGSM and C&W.

Table 5 shows results of the original NIDS mode prior attack. The results are classified under binary class such that 0 represents 'Benign' and 1 represents 'Attack' class. The TN, TP, FN and FP are 48067, 53428, 597 and 125. The area under the ROC curve is 0.9927. The obtained average accuracy, precision, recall and f1-score are 0.9927, 0.9930, 0.9929, and 0.9929, respectively. The evaluation of the model demonstrates outstanding performance in detecting network intrusions while accurately classifying normal network traffic. The NIDS performed optimally under pre-attack cases.

Table 5 NIDS Model Pre-attack Evaluation

| Label | Epsilon (e) / Confidence (c) | TN | TP | FN | FP | ACC | AUC-ROC | P | R | F1 |
|---|---|---|---|---|---|---|---|---|---|---|
| Benign (0) | 0.0000 | 48067 | 53428 | 597 | 125 | 0.9929 | 0.9927 | 0.9974 | 0.9877 | 0.9925 |
| Attack (1) | 0.0000 | | | | | | | 0.9889 | 0.9977 | 0.9933 |
| Weighted Average | | | | | | | | **0.9932** | **0.9927** | **0.9929** |

### *6.2. Performance After Adversarial Attack (Post Attack)*

During the post-attack evaluation phase, the NIDS model’s performance is analyzed after being exposed to adversarial attacks crafted using the FGSM and C&W methods. This evaluation evaluates the vulnerabilities and weaknesses of the NIDS in the presence of adversarial perturbation inputs. By analyzing the NIDS’s response to these attacks, we can identify areas for improvement and develop effective defense mechanisms to enhance its robustness against such threats.

#### *6.2.1. Post FGSM Attack*

We examined a range of epsilon (noise factor $\epsilon$) values in the evaluation to comprehensively evaluate the resilience of the NIDS model. The different epsilon values is motivated by the need to simulate various degrees of FGSM adversarial attacks, from small perturbation noise to high manipulations of network traffic data. By systematically testing these values, we aimed to gain insights into how the NIDS model responds to different levels of adversarial attacks. This method enabled us to examine the model’s robustness and pinpoint the critical stages where its accuracy drops considerably.

Table 6 presents the performance of the NIDS model following the application of the FGSM attack with different epsilon values. Epsilon determines the extent of perturbation introduced to the input data. In this evaluation, epsilon values ranging from 0.0001 to 0.0009 were used to vary the intensity of the FGSM attack. As epsilon increases, the attack becomes more potent, leading to greater perturbation of the input data. The accuracy of the model decreases gradually from 0.8281 to 0.27062. It indicates that the perturbation introduced by the attack makes classification more challenging. Similarly, precision tends to decrease from 0.8543 to 0.2542, indicating that the NIDS becomes more prone to false alarms. The recall value reduces from 0.8346 to 0.2706, indicating that the model misses ‘Attack’ instances and incorrectly classifies these as benign. The F1-score, which balances precision and recall to offer an overall measure of model performance, decreases from 0.8265 to 0.2563 as epsilon values increase.

We further analyzed the model’s robustness using the confusion matrix, focusing on TP, TN, FP and FN. Starting with a small epsilon of 0.0001, our defense correctly identified 37,417 attacks while introducing only 16,136 false positives. This results in a robust F1-Score of 84.32%, demonstrating that our defense effectively distinguishes between malicious and benign traffic. However, as epsilon increases, true positives and false positives decrease. At $\epsilon$ = 0.0002, the model achieves an F1-Score of 65.26%. The number of true positives decreases to 27,066, while false positives rise to 26,487. This pattern continues with higher epsilon values. It leads to a gradual decline in true positives and an increase in false positives. A lower epsilon provides higher sensitivity to attacks but may result in more false positives, while a higher epsilon reduces false alarms but may miss some attacks.

Table 6 Results of post FGSM adversarial attack on NIDS model

| S. No | Epsilon (ϵ) | TN | TP | FN | FP | ACC | AUC-ROC | P | R | F1 |
|---|---|---|---|---|---|---|---|---|---|---|
| 1 | 0.0001 | 47233 | 37417 | 1431 | 16136 | 0.8281 | 0.83464 | 0.8543 | 0.8346 | 0.8265 |
| 2 | 0.0002 | 40231 | 27066 | 8433 | 26487 | 0.6584 | 0.66606 | 0.6827 | 0.6661 | 0.6526 |
| 3 | 0.0003 | 34734 | 18728 | 13930 | 34825 | 0.5230 | 0.53173 | 0.5364 | 0.5317 | 0.5110 |
| 4 | 0.0004 | 28057 | 16613 | 20607 | 36940 | 0.4370 | 0.44338 | 0.4390 | 0.4434 | 0.4299 |
| 5 | 0.0005 | 25654 | 14532 | 23010 | 39021 | 0.3931 | 0.39926 | 0.3919 | 0.3993 | 0.3859 |
| 6 | 0.0006 | 24020 | 12555 | 24644 | 40998 | 0.3578 | 0.36401 | 0.3535 | 0.3640 | 0.3496 |
| 7 | 0.0007 | 22398 | 10846 | 26266 | 42707 | 0. 3252 | 0.33139 | 0.3181 | 0.3314 | 0.3165 |
| 8 | 0.0008 | 20830 | 9309 | 27834 | 44244 | 0. 2949 | 0.30093 | 0.2854 | 0.3009 | 0.2858 |
| 9 | 0.0009 | 19017 | 8057 | 29647 | 45496 | 0.2649 | 0.27062 | 0.2542 | 0.2706 | 0.2563 |

#### *6.2.2. Post C&W Attack*

The same experimentation is conducted on C&W adversarial attacks with varying confidence scores. The confidence parameter provides a means for controlling the adversarial noise of the generated adversarial examples. A higher confidence value leads to more potent adversarial perturbation examples. It makes the adversarial examples complex and increases the likelihood of successful misclassification. The rationale behind exploring different confidence scores is to explore the wide spectrum of adversarial scenarios with varying levels of attacker confidence. Our goal is to deliver a thorough evaluation of the model's performance across various adversarial scenarios, providing key insights to guide decisions on model deployment and potential improvements.

Table 7 depicts the results obtained from the NIDS model after C&W adversarial attack with varying confidence (c) scores. As the confidence score increases from 0.0001 to 0.0009, several notable trends emerge. Firstly, the accuracy decreases progressively from 0.5869 to 0.4961. It indicates that the attack classification is more challenging, reducing overall accuracy. Additionally, precision declines from 0.6376 to 0.5144, indicating an increased likelihood of false alarms. Likewise, recall values decrease, indicating that the model misses some 'Attack' instances and incorrectly classifies as benign. The F1-score, which integrates precision and recall, also declines as the confidence scores increase. These results collectively emphasized the sensitivity of the NIDS model's performance to varying confidence scores in the context of C&W adversarial attacks.

Table 7 Performance post C&W adversarial attack on NIDS models

| S.No | Confidence (c) | TN | TP | FN | FP | ACC | AUC – ROC | P | R | F1 |
|---|---|---|---|---|---|---|---|---|---|---|
| 1 | 0.0001 | 42225 | 17763 | 6439 | 35790 | 0.5869 | 0.59969 | 0.6376 | 0.5997 | 0.5618 |
| 2 | 0.0002 | 40238 | 15982 | 8426 | 37571 | 0.5500 | 0.56264 | 0.5860 | 0.5626 | 0.5232 |
| 3 | 0.0003 | 39772 | 15384 | 8892 | 38169 | 0.5396 | 0.55227 | 0.5720 | 0.5523 | 0.5118 |
| 4 | 0.0004 | 39491 | 14081 | 9173 | 39472 | 0.5241 | 0.53722 | 0.5528 | 0.5372 | 0.4928 |
| 5 | 0.0005 | 39238 | 13543 | 9426 | 40010 | 0.5164 | 0.5296 | 0.5424 | 0.5296 | 0.4837 |
| 6 | 0.0006 | 39043 | 13310 | 9621 | 40243 | 0.5122 | 0.52542 | 0.5364 | 0.5254 | 0.4792 |
| 7 | 0.0007 | 38922 | 13215 | 9742 | 40338 | 0.5101 | 0.52329 | 0.5334 | 0.5233 | 0.4770 |
| 8 | 0.0008 | 38801 | 12324 | 9863 | 41229 | 0.5002 | 0.51373 | 0.5201 | 0.5137 | 0.4642 |
| 9 | 0.0009 | 38694 | 12018 | 9970 | 41535 | 0.4961 | 0.50977 | 0.5144 | 0.5098 | 0.4593 |

Figure 7 shows the AUC-ROC score for both FGSM and C&W adversarial attacks for different epsilon and confidence scores. For the FGSM attack, as the epsilon value increases, indicating stronger attacks with larger perturbations to the input data, the model's AUC-ROC score steadily decreases. This behaviour signifies that FGSM attacks progressively compromise the model's capability to differentiate between benign and malicious traffic. At lower epsilon values, the model exhibits robust discrimination, while at higher epsilon values, it struggles to differentiate effectively.

In contrast, the C&W attack, which used confidence scores, shows different behaviour. The model maintains reasonable discrimination between benign and attack traffic at lower confidence levels, as reflected by higher AUC-ROC scores. However, as the confidence level increases, indicating greater attacker confidence in generating adversarial examples, the model's discrimination power declines. The AUC-ROC scores gradually decrease as confidence levels rise. It signifies that C&W attacks with higher confidence challenge the model's ability to classify traffic accurately.

In summary, while both FGSM and C&W attacks impact model performance negatively as they become stronger, FGSM's behavior is primarily determined by the magnitude of the attack (epsilon), whereas C&W's behavior is influenced by the attacker's confidence in crafting adversarial examples. Researchers should consider these distinct behaviours when evaluating the robustness of their models against different attack types.

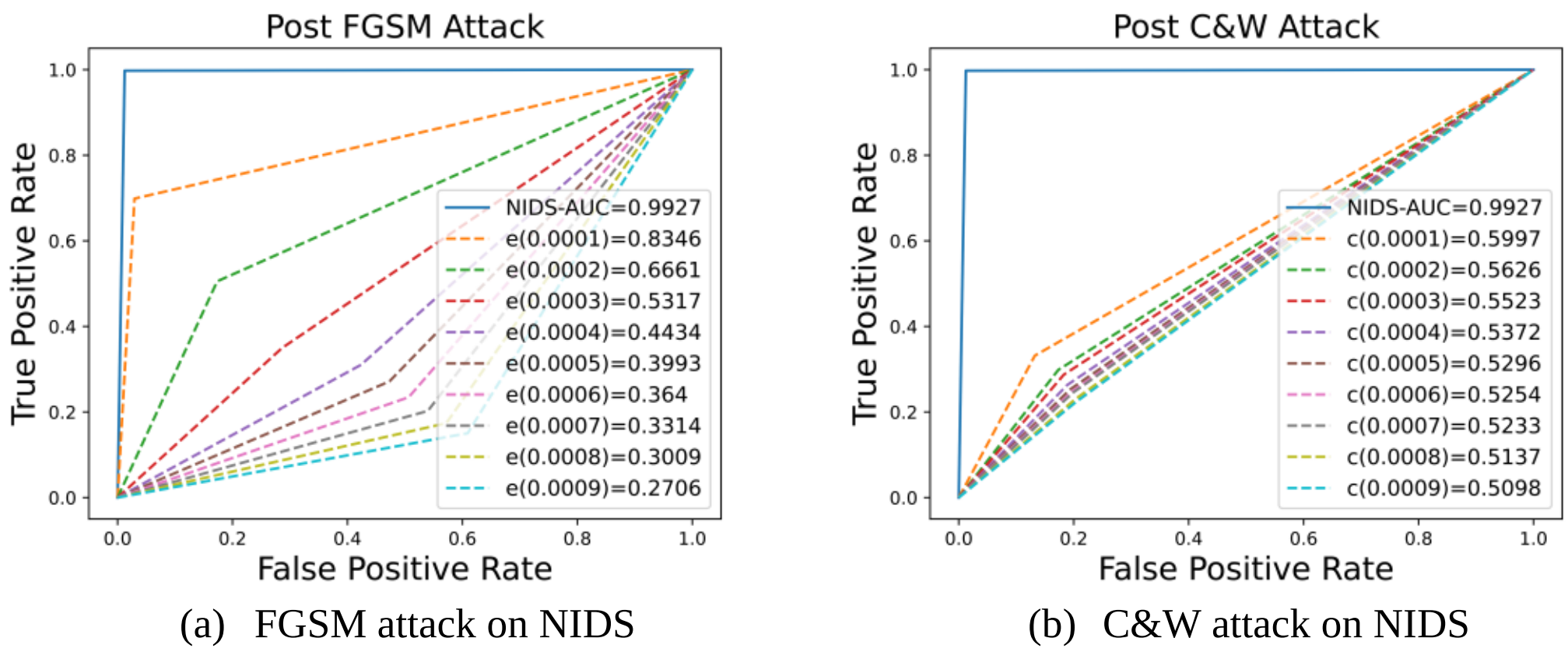

(a) FGSM attack on NIDS (b) C&W attack on NIDS

Figure 7 (a) AUC-ROC score after applying FGSM attack (b) AUC-ROC score after applying C&W attack

### *6.3. Performance After Hybrid Defense (Post Defense)*

#### *6.3.1. Post Defense Evaluation on FGSM*

The proposed defense helps the NIDS model maintain a higher level of accuracy and f1-score which are crucial for correctly identifying and classifying network intrusions and attacks. The defense method acts as a countermeasure against adversarial attacks, making the NIDS more resilient.

Table 8 shows the different results after applying the post defense method over the NIDS model after FGSM adversarial attacks execution. For example, with an epsilon of 0.0009, accuracy increased to 0.9657, indicating a recovery in overall classification performance. Precision improved to 0.9677, suggesting a reduction in false alarms or false 'Attack' classifications. Recall improved to 0.9644, indicating fewer 'Attack' instances were missed, leading to fewer false negatives. F1-score improved to 0.9655, reflecting an overall enhancement in NIDS model effectiveness. The other result we can see with an epsilon of 0.0005, accuracy is slightly higher, i.e., 0.9695 compared to epsilon=0.0009. Similarly, the precision improved to 0.9715. The recall and f1-score are also increased to 0.9683, 0.9694. The overall results show the enhancement in NIDS model robustness. The highest accuracy, precision, recall, f1-score and roc-auc obtained for e =0.0001 are 0.9723, 0.9741, 0.9712, 0.9722 and 0.9712.

It is noteworthy that there is a substantial improvement in performance even for all values of epsilon. This statement emphasizes the significant enhancement in model performance across all epsilon values, highlighting the performance of the post-defense method to protect against FGSM attacks.

#### *6.3.2. Post Defense Evaluation on C&W*

Table 9 shows the proposed defense results on the NIDS model. Improvements were observed for different confidence scores ranging from 0.0001 to 0.0009. For example, epsilon (e) = 0.0001 the values of TN, TP, FN, FP, accuracy, AUC-ROC, precision, recall, f1-score 41066, 53258, 7598, 295, 0.9228, 0.9192, 0.9929, 0.8439 and 0.9123, respectively. These results indicate a high accuracy and precision with low FP and FN for adversarial attacks at a confidence score of 0.0001 after applying the post-defense method. For epsilon= 0.0002, results show a slightly lower accuracy and precision compared to the previous scenario. However, the model performs well with precision and recall. For epsilon 0.0003 - 0.0009, as epsilon increases, there is a gradual decline in model performance. However, the model is still able to obtain a good accuracy of 0.88688, even with the highest confidence value of 0.0009.

Table 8 Results of post defense under FGSM adversarial attack

| S.No | Epsilon (e) | TN | TP | FN | FP | ACC | AUC-ROC | P | R | F1 |
|---|---|---|---|---|---|---|---|---|---|---|
| 1 | 0.0001 | 46125 | 53262 | 2539 | 291 | 0.9723 | 0.9712 | 0.9741 | 0.9712 | 0.9722 |
| 2 | 0.0002 | 46077 | 53253 | 2587 | 300 | 0.9718 | 0.97062 | 0.9736 | 0.9706 | 0.9716 |
| 3 | 0.0003 | 46031 | 53243 | 2633 | 310 | 0.9712 | 0.97005 | 0.9731 | 0.9701 | 0.9711 |
| 4 | 0.0004 | 45953 | 53227 | 2711 | 326 | 0.9703 | 0.9691 | 0.9722 | 0.9691 | 0.9701 |
| 5 | 0.0005 | 45888 | 53214 | 2776 | 339 | 0. .9695 | 0.96831 | 0.9715 | 0.9683 | 0.9694 |
| 6 | 0.0006 | 45855 | 53194 | 2809 | 359 | 0.9690 | 0.96779 | 0.9710 | 0.9678 | 0.9688 |
| 7 | 0.0007 | 45813 | 53165 | 2851 | 388 | 0.9683 | 0.96708 | 0.9704 | 0.9671 | 0.9681 |
| 8 | 0.0008 | 45709 | 53102 | 2955 | 451 | 0.9667 | 0.96543 | 0.9688 | 0.9654 | 0.9665 |
| 9 | 0.0009 | 45681 | 53027 | 2983 | 526 | 0.9657 | 0.96444 | 0.9677 | 0.9644 | 0.9655 |

Furthermore, in scenario 1 ($\epsilon = 0.0001$), we observed 53258 true positives (correctly classified attacks), 41066 true negatives 295 false positives (benign traffic misclassified as attacks), and 7598 false negatives (attacks misclassified as benign). Similar trends continue in the subsequent scenarios with varying epsilon values, each reflecting the NIDS's ability to correctly classify attacks and benign traffic, as well as its susceptibility to false positives and false negatives. These results demonstrate the model's resilience and performance in the presence of adversarial threats with increasing epsilon values. The post-defense method is effective in improving the model's performance under C&W attacks with low confidence scores (e.g., 0.0001). It achieved high accuracy and precision. However, with the increasing value of confidence, the decrease in performance is not that high. It shows that the post-defense method is relatively effective even for higher confidence scores.

Table 9 Results of post defense under C&W adversarial attack

| S.No | Confidence (e) | TN | TP | FN | FP | ACC | AUC-ROC | P | R | F1 |
|---|---|---|---|---|---|---|---|---|---|---|
| 1 | 0.0001 | 41066 | 53258 | 7598 | 295 | 0.9228 | 0.91918 | 0.9340 | 0.9192 | 0.9217 |
| 2 | 0.0002 | 39260 | 53273 | 9404 | 280 | 0.9053 | 0.90076 | 0.9214 | 0.9008 | 0.9034 |
| 3 | 0.0003 | 38950 | 53260 | 9714 | 293 | 0. 9021 | 0.89746 | 0.9191 | 0.8975 | 0.9001 |
| 4 | 0.0004 | 38614 | 53273 | 10050 | 280 | 0.8989 | 0.89413 | 0.9170 | 0.8941 | 0.8968 |
| 5 | 0.0005 | 38523 | 53258 | 10141 | 295 | 0.8979 | 0.89305 | 0.9162 | 0.8931 | 0.8957 |

| 6 | 0.0006 | 38309 | 53261 | 10355 | 292 | 0.8958 | 0.89088 | 0.9148 | 0.8909 | 0.8936 |
|---|---|---|---|---|---|---|---|---|---|---|
| 7 | 0.0007 | 38221 | 53255 | 10443 | 298 | 0.8949 | 0.88992 | 0.9142 | 0.8899 | 0.8926 |
| 8 | 0.0008 | 38027 | 53265 | 10637 | 288 | 0.8931 | 0.88802 | 0.9130 | 0.8880 | 0.8907 |
| 9 | 0.0009 | 37912 | 53269 | 10752 | 284 | 0. 8920 | 0.88688 | 0.9123 | 0.8869 | 0.8895 |

Figure 8 (a) and Figure 8 (b) show the AUC-ROC score for different epsilon and confidence values. The AUC-ROC scores for the FGSM and C&W attacks after applying the proposed method indicate their different behaviours. The analysis of the proposed defense method over AUC-ROC shows that the proposed defense mitigated the attack's impact in both FGSM and C&W attacks. The AUC-ROC scores for initial epsilon and confidence values of 0.0001, for FGSM and C&W attacks are 0.9712 and 0.9192. Gradually, when both noise factor increase from 0.0001 to 0.0009, the AUC-ROC scores for C&W also slightly decrease, although the decrease is not as significant as seen with C&W. The results indicate that the C&W is a more potent attack compared to FGSM.

In summary, the proposed method shows a more pronounced impact on mitigating the FGSM attack. While it also affects the C&W attack, the reduction in AUC-ROC scores is high compared to C&W. It indicates the defense method is more effective on FGSM compared to C&W.

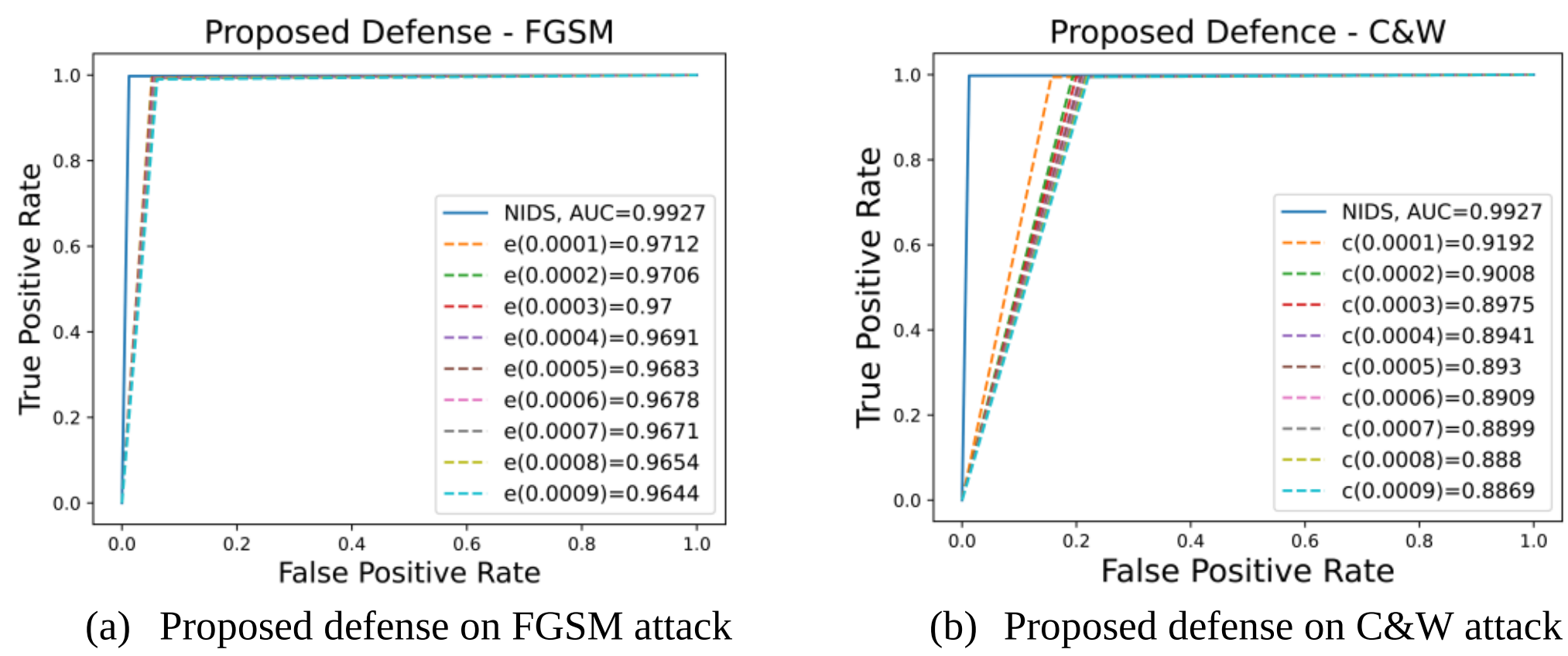

(a) Proposed defense on FGSM attack (b) Proposed defense on C&W attack

Figure 8 (a) AUC-ROC score after applying proposed defense on FGSM attack (b) AUC-ROC score after applying proposed defense on C&W attack

*6.4. Results Interpretation and Comparison*

Table 10, along with Figure 9, shows the quantification of the final results. We compared performance before and after applying the hybrid method. A noticeable degradation in model performance is observed under both attack types. Specifically, FGSM causes a sharp drop in accuracy from 82.81% to 26.49% and in F1-score from 82.65% to 25.63% as $\epsilon$ increases from 0.0001 to 0.0009. The C&W attack also leads to performance degradation, with accuracy reducing from 58.69% to 49.61% and F1-score from 56.18% to 45.93%. However, the hybrid defense effectively mitigates these effects. It maintained robust classification performance across all perturbation levels. Under the FGSM attack, accuracy improved from 26.49% to 96.57%, and under the C&W attack, from 49.61% to 89.20%. This stability confirms the resilience of the model against varying levels of perturbation.

Figure 9 provides a demographic representation of the same results. It clearly highlights the gap between attacked and defended performance. The figure effectively demonstrates how the hybrid defense brings the model's performance close to its original, non-attacked state. In summary, both the tabular and graphical results validate that while the NIDS is vulnerable to adversarial noise, the proposed hybrid defense strategy offers strong resilience and ensures reliable detection even under increasing attack intensity.

Table 10 Consolidated accuracy and F1 score under FGSM and C&W attacks with and without hybrid defense

| S. No | Epsilon (ϵ) | Without Hybrid Defense | | With Hybrid Defense | | Confidence (c) | Without Hybrid Defense | | With Hybrid Defense | |
|---|---|---|---|---|---|---|---|---|---|---|
| | | Post FGSM Attack ACC | Post FGSM Attack F1 | Post Hybrid Defense ACC | Post Hybrid Defense F1 | | Post C&W Attack ACC | Post C&W Attack F1 | Post Hybrid Defense ACC | Post Hybrid Defense F1 |
| 1 | 0.0001 | 0.8281 | 0.8265 | 0.9723 | 0.9722 | 0.0001 | 0.5869 | 0.5618 | 0.9228 | 0.9217 |
| 2 | 0.0002 | 0.6584 | 0.6526 | 0.9718 | 0.9716 | 0.0002 | 0.55 | 0.5232 | 0.9053 | 0.9034 |
| 3 | 0.0003 | 0.523 | 0.511 | 0.9712 | 0.9711 | 0.0003 | 0.5396 | 0.5118 | 0. 9021 | 0.9001 |
| 4 | 0.0004 | 0.437 | 0.4299 | 0.9703 | 0.9701 | 0.0004 | 0.5241 | 0.4928 | 0.8989 | 0.8968 |
| 5 | 0.0005 | 0.3931 | 0.3859 | 0. .9695 | 0.9694 | 0.0005 | 0.5164 | 0.4837 | 0.8979 | 0.8957 |
| 6 | 0.0006 | 0.3578 | 0.3496 | 0.969 | 0.9688 | 0.0006 | 0.5122 | 0.4792 | 0.8958 | 0.8936 |
| 7 | 0.0007 | 0. 3252 | 0.3165 | 0.9683 | 0.9681 | 0.0007 | 0.5101 | 0.477 | 0.8949 | 0.8926 |
| 8 | 0.0008 | 0. 2949 | 0.2858 | 0.9667 | 0.9665 | 0.0008 | 0.5002 | 0.4642 | 0.8931 | 0.8907 |
| 9 | 0.0009 | 0.2649 | 0.2563 | 0.9657 | 0.9655 | 0.0009 | 0.4961 | 0.4593 | 0. 8920 | 0.8895 |

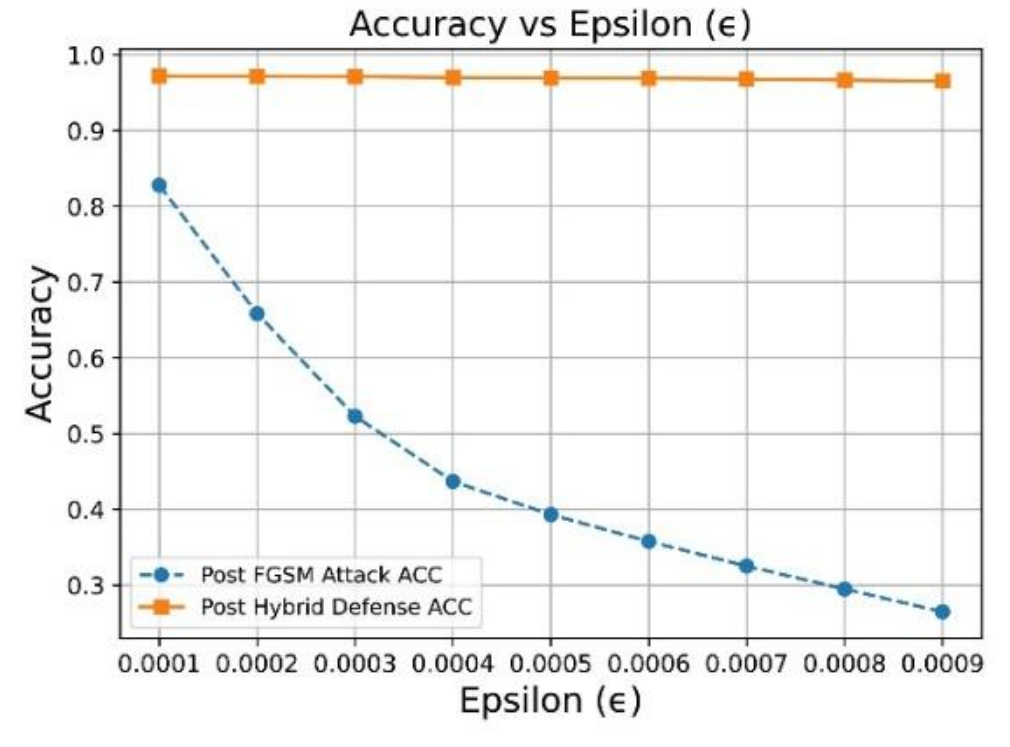

(a) Accuracy for FGSM with or without hybrid defense

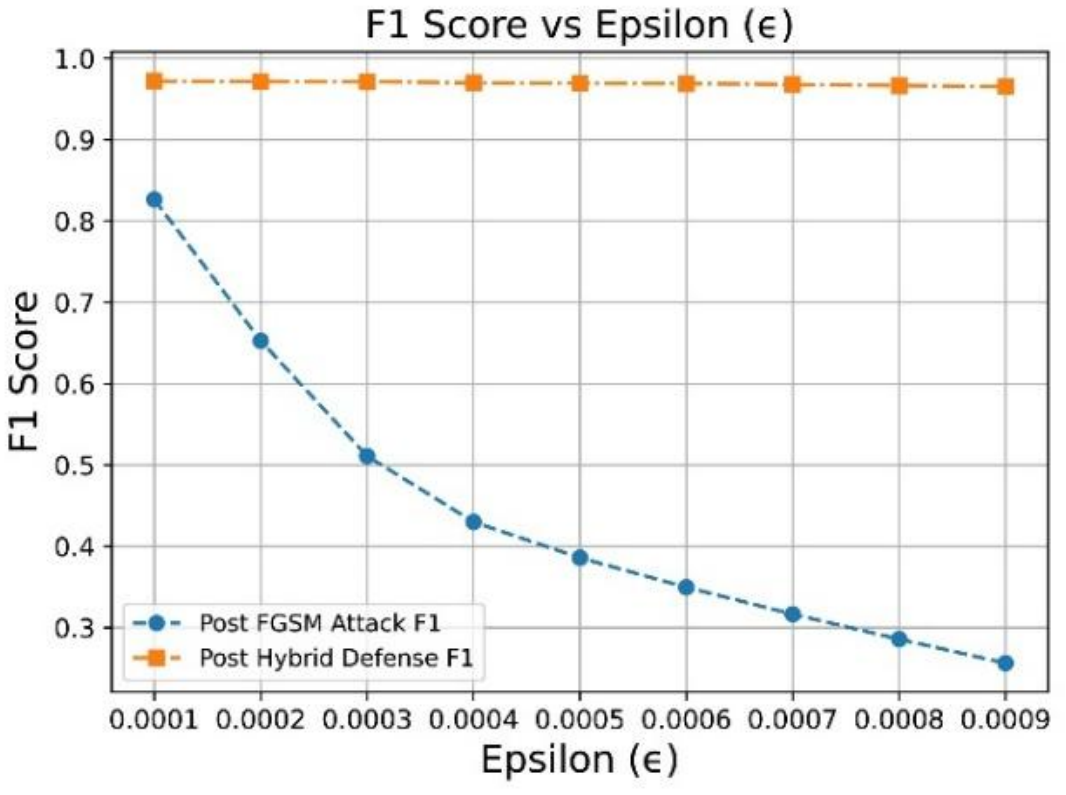

(b) F1 Score for FGSM with or without hybrid defense

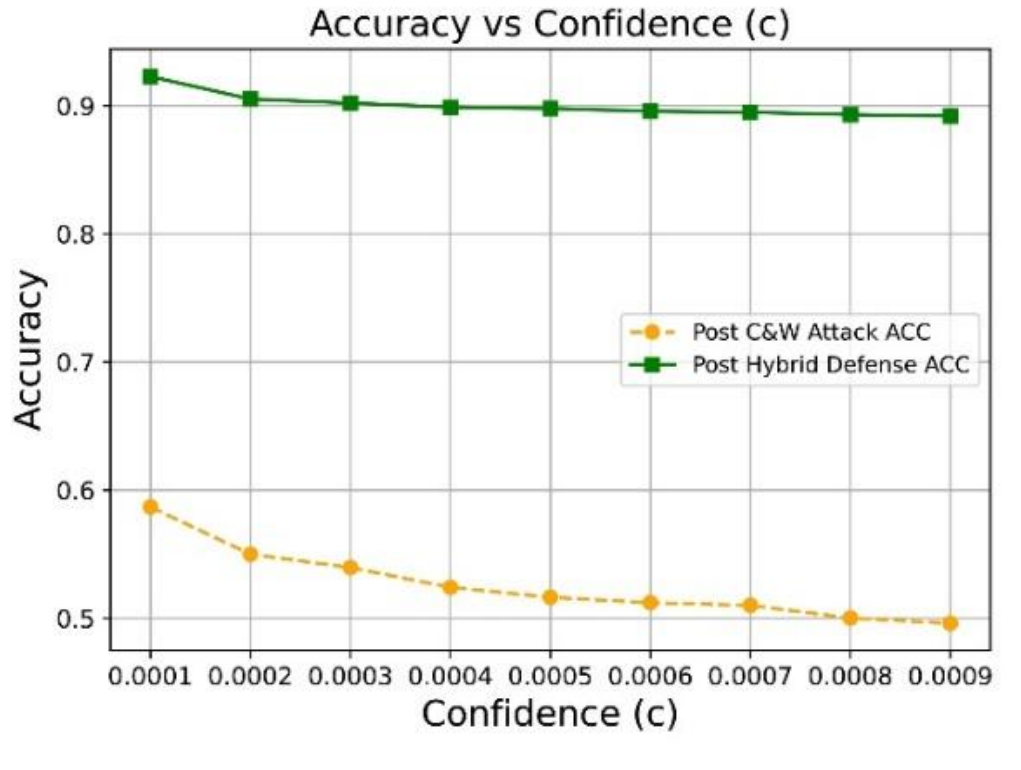

(c) Accuracy for C&W with or without hybrid defense

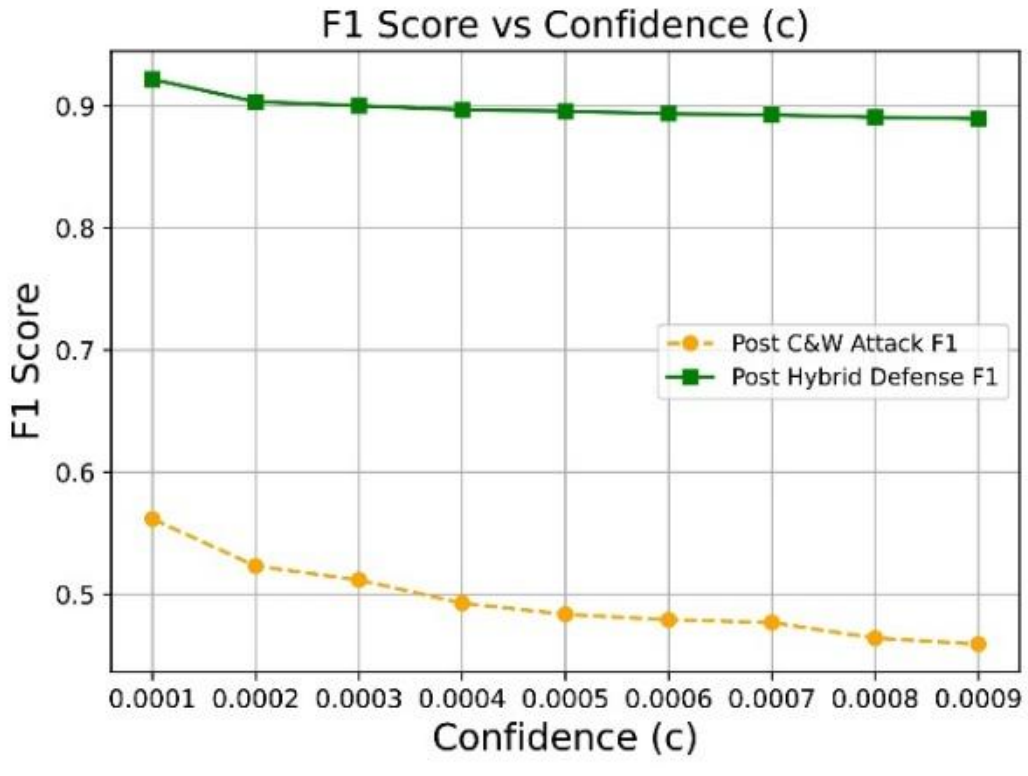

(d) F1 Score for C&W with or without hybrid defense

Figure 9 Demographic representation of Accuracy and F1 score under FGSM and C&W attacks with and without hybrid defense

Table 11 provided a comparative analysis of the proposed hybrid defense model against existing state-of-the-art techniques in terms of accuracy on benchmark datasets under adversarial conditions. The proposed model achieved a high pre-attack accuracy of 99.29%. However, accuracy dropped significantly to 49.46% and 52.61%, respectively, under adversarial attack. Importantly, the proposed hybrid defense recovered accuracy to 96.95% against FGSM and 90.12% against C&W. It showed the strong resilience and generalization.

Roshan and Zafar [29] used a deep learning model with GDA and feature squeezing on CICIDS-2017. The proposed method achieved post-defense accuracy of 89.7% and 89.5% against FGSM and C&W, respectively. Debicha et al. [33] reported 74.50% accuracy using a transfer learning-based adversarial defense across the NSL-KDD and CICIDS-2017 datasets, while Hashemi et al. [48] achieved only a 49% detection rate through re-training with partial observations. Usama et al. [41] proposed a GAN-based retraining approach to enhance the robustness of a deep neural network (DNN) on the KDDCUP-99 dataset. The model initially achieved an F1-score of 80.03% before the attack. However, under adversarial conditions, the F1-score dropped significantly to 40.86%. After applying adversarial training with GAN-generated samples, the post-defense F1-score improved to 83.56%.

Our proposed defense strategy outperformed existing methods in terms of post-attack recovery and achieved near-optimal detection rates even under strong adversarial scenarios. It validates its effectiveness and practicality for real-world intrusion detection systems.

Table 11 Comparison of the proposed model with state-of-the-art Research

| Reference | Model | Attack | Defense | Dataset | Accuracy (Average) |
|---|---|---|---|---|---|
| **Proposed Defense** | DL Model | FGSM and C&W | Hybrid Defense Method (GDA and AT) | CICIDS-2017 | **Pre-Attack: 99.29%**<br>**Post Attack: 49.46% (FGSM), 52.61%(C&W)**<br>**Post Hybrid Defense: 96.95% (FGSM), 90.12% (C&W).** |

| Roshan and Zafar [29] (2024) | DL Model | FGSM, C&W | GDA and Feature Squeezing | CICIDS-2017 | Post Defense:<br>89.7 % and 89.5% (Accuracy<br>) |
|---|---|---|---|---|---|
| Debicha et al. [33] (2023) | DL Model | FGSM, PGD, DeepFool | Transfer Learning Defense | NSL-KDD,<br>CICIDS-2017 | Post Defense :<br>74.50 % (Accuracy),<br>83.67% (Accuracy) |
| Hashemi et al. [48] (2020) | Partial Observation on Reconstruction | Hashemi | Re-Training | CICIDS-2017 | 49 % (Detection Rate) |
| Usama et al. [41] (2019) | DNN, GAN | KDDCUP-99 | GAN based Re Training | Adversarial Training | Pre Attack:<br>80.03 % (F1 Score)<br>Post Attack:<br>40.86 % (F1 Score)<br>Post Defense: F: 83.56 % (F1 Score) |

## 7. Limitations and Future Scope

Some of the constraints and potential research directions for new researchers interested in cybersecurity and adversarial machine learning.

- Adversarial Attacks: The research investigated two widely known adversarial attack methods, FGSM and C&W. However, adversarial threats is rapidly evolving with diverse adversarial attack scenarios such as PGD, JSMA, BIM, etc. Hence, confining the research with a particular attack may not fully encompass the wide range of real-world situations.
- Testbed Dataset: This study uses the CICIDS-2017 dataset for evaluation. However, the static nature of the dataset may not accurately reflect the dynamic and constantly changing characteristics of network traffic and adversarial evasion tactics. Hence, the conclusions drawn from the study may be limited to only static evaluation and not entirely align with real-world operational settings. Additional datasets may be used to explore the real-world scenarios further within NIDS.
- Black-box Adversarial Attack: The proposed study concentrates on the white-box attack approach. Here the adversary has comprehensive information of the targeted model for exploitation [43][[44]. However, it does not represent the real-world scenario as, in most cases, the attacker/hacker does not have information about the targeted system.

**Future Scope of the Study:**

Adversarial machine learning is a rapidly evolving field which offers numerous exciting avenues for future research and development.

- New Adversarial Attacks: The researcher should investigate the impact of new adversarial techniques such as grey-box attacks etc. This will help enhance the resilience and robustness of NIDS against various attack scenarios.

- Black-Box Transferability Attack: In the future, researchers may explore black-box attacks to overcome the weaknesses of basic model attacks. It also refers to the black-box transferability attacks. The adversary uses the concept of surrogate model to mimic the behavior of the targeted system.
- Exploration of Explainability: Explainability is an emerging research area. Hence, incorporating clear explanations and user-friendly information into NIDS models is a promising concept. When these models can provide straightforward explanations for their predictions, it greatly assists security administrators in quickly identifying and responding to cyberattacks.
- Real-World Deployment: Real-world implementation is essential to authenticate and validate NIDS defense. Practical implementation in live/simulated network environments will provide real-time performance and the practical applicability of NIDS and defensive methods [49][50].

### 8. Conclusion

The proposed research rigorously investigates the vulnerabilities of NIDS against evolving adversarial attacks. The critical vulnerability is exposed when evaluating NIDS to adversarial attacks, specifically FGSM and C&W. These attacks significantly compromised the NIDS performance such as classification report and confusion matrix, particularly when the adversarial noise intensity increases from 0.0001 to 0.0009 (epsilon and confidence values). The proposed defense method combines two heuristic defense methods: Adversarial training (AT) and Gaussian data augmentation (GDA), which uses the stacking integration method and the concept of metaclassifier. The post-defense evaluation of the NIDS demonstrates the improvement in its performance and indicates its robustness when encountered against adversarial attacks. The accuracy and f1-score of post-defense evaluation are (0.9657, 0.9655) (0.8920, 0.8895) for 0.0009 noise factor.

Additionally, the proposed defense algorithm has applications beyond network security and could be explored in various fields such as video analytics, image processing, healthcare systems, and more, demonstrating its versatility and adaptability. Furthermore, integrating adversarial transferability concepts is also a great direction to be explored.

## Declarations and Statements

**Conflicts of Interest:** The authors declare no conflict of interest.

**Code and Dataset Availability Statement:** The code developed during this study will be provided upon request. **Dataset link:** https://www.unb.ca/cic/about/index.html

## Abbreviations

### Datasets & Systems

**NIDS** – Network Intrusion Detection System

**RAIDS** – Robust Autoencoder-based Intrusion Detection System

**CTU-13** – Czech Technical University 2013 Botnet Dataset

**CICIDS-2017** – Canadian Institute for Cybersecurity Intrusion Detection System 2017 Dataset

**NSL-KDD** – Network Security Laboratory - Knowledge Discovery in Databases

**UNSW-NB-15** – University of New South Wales Network-Based 2015 Dataset

**AWID** – Aegean Wireless Intrusion Detection Dataset

### Models & Algorithms

**ANN** – Artificial Neural Network

**CNN** – Convolutional Neural Network

**RNN** – Recurrent Neural Network

**LSTM** – Long Short-Term Memory

**DAE** – Denoising Autoencoder

**SVM** – Support Vector Machine

**KNN** – K-Nearest Neighbours

**RF** – Random Forest

**ADB** – AdaBoost

**QDA** – Quadratic Discriminant Analysis

**GNB** – Gaussian Naive Bayes

**GAN** – Generative Adversarial Network

**GA** – Genetic Algorithm

**PSO** – Particle Swarm Optimization

**<u>Adversarial Attacks</u>**

**FGSM** – Fast Gradient Sign Method

**BIM** – Basic Iterative Method

**PGD** – Projected Gradient Descent

**C&W** – Carlini and Wagner Attack

**JSMA** – Jacobian-Based Saliency Map Attack

**<u>Evaluation Metrics & Others</u>**

**A** – Accuracy

**F** – F1-Score

**P** – Precision

**R** – Recall

**FPR** – False Positive Rate

**AUC** – Area Under ROC Curve

**ASR** – Attack Success Rate

**DER** – Detection Evasion Rate

**ER** – Evasion Rate

**MER** – Malicious Traffic Evasion Rate

**MMR** – Malicious Features Mimicry Rate

**PDR** – Probability Decline Rate

**MAPE** – Mean Absolute Percentage Error

**DT** – Detection Rate

**Attack Settings**

**BB** – Black-box

**GB** – Grey-box

**WB** – White-box